\documentclass[aps,pra,reprint,twocolumn,superscriptaddress,showpacs,showkeys,superscriptaddress,longbibliography]{revtex4-1}
\usepackage{amsmath,amssymb,amstext}
\usepackage[usenames,dvipsnames]{color}
\usepackage{graphicx}
\usepackage{enumitem}
\usepackage{bm,bbold,braket}
\usepackage{natbib}
\usepackage{txfonts, comment,stmaryrd}
\usepackage{dcolumn}
\usepackage{color}
\usepackage{xcolor}
\colorlet{rn}{red}
\colorlet{an}{blue}
\usepackage[english]{babel}

\SetLabelAlign{Center}{\hfil#1\hfil}
\SetLabelAlign{CenterWithParen}{\hfil(\makebox[1.0em]{#1})\hfil}

\usepackage[colorlinks,bookmarks=false,citecolor=blue,linkcolor=red,urlcolor=blue]{hyperref}

\begin{document}

\title{Patterns, spin-spin correlations and competing instabilities in driven quasi-two-dimensional spin-1 Bose-Einstein condensates}
\author{Sandra M Jose}
\affiliation{Department of Physics, Indian Institute of Science Education and Research, Pune 411 008, India}  
\author{Komal Sah}
\affiliation{Department of Physics, Indian Institute of Science Education and Research, Pune 411 008, India}   
\affiliation{Department of Physics, University of California, Davis 95616, USA}   
\author{Rejish Nath}
\affiliation{Department of Physics, Indian Institute of Science Education and Research, Pune 411 008, India}    


\begin{abstract}
We analyze the formation of transient patterns and spin-spin correlations in quasi-two-dimensional spin-1 homogeneous Bose-Einstein condensates subjected to parametric driving of $s$-wave scattering lengths. The dynamics for an initial ferromagnetic phase is identical to that of a scalar condensate. In contrast, intriguing dynamics emerges for an initial polar state. For instance, we show that competition exists between density patterns and spin-mixing dynamics. Dominant spin-mixing dynamics lead to a gas of polar core vortices and anti-vortices of different spin textures. The density modes of the Bogoliubov spectrum govern the wavenumber selection of Faraday patterns. The spin modes determine the vortex density and the spatial dependence of spin-spin correlation functions. When the density patterns outgrow the spin-mixing dynamics, the spin-spin correlations decay exponentially with a correlation length of the order a spin healing length; otherwise, they exhibit a Bessel function dependence. Strikingly, competing instabilities within density and spin modes emerge when both scattering lengths are modulated at different frequencies and appropriate modulation amplitudes. The competing instability leads to a superposition of density patterns or correlation functions of two distinct wavelengths. Our studies reveal that fine control over the driven dynamics can be attained by tuning interaction strengths, quadratic Zeeman field, driving frequencies, and amplitudes. 
\end{abstract}

\maketitle

\section{Introduction}
%
Periodically driven Bose-Einstein condensates (BECs) have been a playground for studying various phenomena such as Faraday patterns \cite{sta02, eng07, nic07, nat10,sta04,mod06,kat10, nic11,cap11,kaz12,bal12,bal14,tur20,bha08,kaz12,sud16, com22, zha22}, dark-soliton lattice \cite{ver17}, dynamical localization \cite{lig07,eck05,zen09}, spin freezing \cite{zha10, hoa13}, matter-wave jets \cite{cla17, han18, zhi19,han20}, bright solitons \cite{sai03,abd03}, parametric instability \cite{lel17}, generating higher harmonics \cite{ber23} etc. Faraday patterns are observed experimentally in elongated condensates by modulating either the trap frequencies \cite{eng07, com22} or the interactions \cite{ngu19,zha20}. Such patterns offer critical insights into the elementary excitations of condensates because the pattern size is determined by the Bogoliubov mode resonant with half of the driving frequency. A non-monotonous excitation spectrum can make the wavenumber selection non-trivial \cite{nat10, kaz12}. 


Because of the spin degrees of freedom, spinor condensates are ideal for exploring spin textures and magnetic phases \cite{sad06, sta13, kaw12}. Interesting phenomena such as dynamical stabilization \cite{hoa13}, spin-squeezing \cite{sai08, hoa16}, Shapiro resonances \cite{evr19, ima21}, parametric resonances \cite{pen21} and the quantum walk in momentum space \cite{dad18, dad19} have been reported in driven spinor condensates. In this paper, we analyze the formation of transient density and spin patterns, and spin-spin correlations in a quasi-two-dimensional (Q2D) spin-1 homogeneous condensate subjected to the parametric driving of $s$-wave scattering lengths $a_0$ and $a_2$. We consider three cases where either $a_0$ or $a_2$ is modulated individually, and both are modulated simultaneously. Modulating either $a_0$ or $a_2$ leads to modulations in both spin-independent and spin-dependent interactions, affecting dynamics of both spatial and spin degrees of freedom.

As we show, the dynamics depends critically on the initial states and in particular, we consider ferromagnetic and polar phases. A ferromagnetic phase is immune to the modulation of $a_0$ \cite{Kom19}, whereas $a_2$ modulation results in dynamics similar to that of a driven scalar condensate \cite{sta02, eng07}. In contrast, an initial polar condensate exhibits non-trivial dynamics. Unstable Bogoliubov modes lead to both Faraday patterns and spin-mixing dynamics. There exists an implicit competition between density modulations and spin-mixing dynamics. When the spin-mixing becomes dominant, a gas of polar core vortices (PCVs) and anti-vortices of different spin textures are formed. Previous studies on PCVs in Q2D spin-1 condensates are based on the Kibble-Zurek mechanism via quenching the quadratic Zeeman field from non-polar phases \cite{sai08,sai07a,wil16}. The unstable momentum of the density mode quantifies the wavenumber of the Faraday pattern, whereas that of the spin mode determines the vortex density. At longer times, spin-mixing disrupts the selection of higher harmonics, which is in high contrast to the case of a scalar condensate. If the spin-dependent interactions are stronger than the spin-independent ones, the Faraday patterns outgrow the spin dynamics, and the spin-spin correlations decay exponentially over distance with a correlation length of the order a spin healing length. In contrast, when the spin dynamics are dominant, the spin-spin correlations are governed by a Bessel function with an argument depending on the momentum of the unstable spin mode.

Interestingly, when both scattering lengths are modulated simultaneously with the same frequency, the modulation amplitudes can be chosen such that parametric driving is present only in spin-independent or spin-dependent interactions. Hence, it is possible to excite the density or spin mode alone during the initial stage of the dynamics. When the modulation frequencies differ, a fascinating scenario of competing instabilities appears. For instance, two momenta from either density or spin mode become equally unstable, causing competition between the two wavelengths. In a classical fluid, two patterns with different symmetries coexist in the onset of competing instabilities \cite{kum95}. In our case, the competing instability results in a superposition of Faraday patterns or correlation function with different wavelengths whose amplitudes vary in time. To conclude, our studies reveal that the nature of driven dynamics in spinor condensates can be controlled by tuning the interaction strengths, quadratic Zeeman field, modulation frequencies, and amplitudes.

The paper is structured as follows. In Sec.~\ref{sm}, we discuss the setup and the meanfield equations describing a spin-1 condensate. The Mathieu-like equations governing the wavenumber selection of a driven spin-1 condensate is derived in Sec.~\ref{tm}. The dynamics of driven ferromagnetic and polar phases, including the spin-spin correlations is discussed in Sec.~\ref{fmp} and Sec.~\ref{pp}, respectively. Competing instability of different density or spin modes is discussed in Sec.~\ref{ci}. The experimental possibilities are discussed in Sec.~\ref{exp}. Finally, we summarize and provide an outlook in Sec.~\ref{sum}.


\section{Setup and Model}
\label{sm}

We consider a Q2D spin-1 homogeneous Bose gas in the presence of a quadratic Zeeman field $q$. The Hamiltonian describing the system is 
\begin{align}
\hat H=\int d{\bm \rho}\sum_{m=0, \pm 1}\hat\psi_m^{\dagger}({\bm \rho})\left(-\frac{\hbar^2}{2M}\nabla_{\rho}^2+qm^2\right)\hat\psi_m(\bm \rho)+\hat H_Z+\hat V_I,
\label{h1}
\end{align}
where $\hat\psi_m$ is the field operator which annihilates a boson in the $m$th Zeeman state, $M$ is the mass of a boson and ${\bm \rho}=(x, y)$. The quadratic Zeeman Hamiltonian is, 
\begin{equation}
\hat H_Z=q\int d{\bm \rho}\sum_{m_1, m_2}\hat\psi_{m_1}^{\dagger}({\bm \rho})\left(\hat F_z^2\right)_{m_1, m_2}\hat\psi_{m_2}({\bm \rho}),
\label{hz}
\end{equation}
and the interaction operator is
\begin{equation}
\hat V_I=\frac{1}{2}\int d{\bm \rho}\left[\tilde c_0:\hat n^2({\bm \rho}):+\tilde c_1:\hat {\bm F}^2({\bm \rho}):\right],
\label{hi}
\end{equation}
where $\tilde c_{0, 1}=c_{0, 1}/\sqrt{2\pi}l_z$ with $l_z=\sqrt{\hbar/m\omega_z}$ being the transverse width of the condensate provided by the harmonic potential $V_t(z)=m\omega_z^2z^2/2$, $\hat n({\bm \rho})=\sum_{m=-f}^f\hat\psi_m^{\dagger}({\bm \rho})\hat\psi_m({\bm \rho})$ is the total density operator. The symbol : : denotes the normal ordering that places annihilation operators to the right of the creation operators. The components of the spin density operator are
\begin{equation}
\hat {\bm F}_{\nu\in x, y, z}({\bm \rho})=\sum_{m, m'}\left(f_\nu\right)_{mm'}\hat\psi_m^{\dagger}({\bm \rho})\hat\psi_{m'}({\bm \rho}),
\end{equation}
with $f_\nu$ being the $\nu$th component of the spin-1 matrices. The spin-independent and spin-dependent interaction constants are $c_0=(g_0+2g_2)/3>0$ and $c_1=(g_2-g_0)/3$, respectively with $g_{\mathcal F}=4\pi\hbar^2a_{\mathcal F}/m$ related to the scattering length $a_{\mathcal F=0, \ 2}$ of the total spin-${\mathcal F}$ channel. 

At very low temperatures the system is described by the coupled non-linear Gross-Pitaevskii equations (NLGPEs), where $\hat\psi_m({\bm \rho})$ is replaced by a c-number $\psi_m({\bm \rho})$,  
\begin{eqnarray}
\label{m1}
i\hbar\frac{\partial \psi_{1}}{\partial t}&=&\left[-\frac{\hbar^2\nabla_\rho^2}{2M}+q+\tilde c_0n+\tilde c_1 F_z\right]\psi_1+\frac{\tilde c_1F_-}{\sqrt{2}} \psi_0 \\
\label{m2}
i\hbar\frac{\partial \psi_{0}}{\partial t}&=&\left[-\frac{\hbar^2\nabla_\rho^2}{2M}+\tilde c_0n\right]\psi_0+\frac{\tilde c_1}{\sqrt{2}} F_+\psi_1+\frac{\tilde c_1F_-}{\sqrt{2}}\psi_{-1} \\
\label{m3}
i\hbar\frac{\partial \psi_{-1}}{\partial t}&=&\left[-\frac{\hbar^2\nabla_\rho^2}{2M}+q+\tilde c_0n-\tilde c_1F_z\right]\psi_{-1}+\frac{\tilde c_1F_+}{\sqrt{2}}\psi_{0},
\end{eqnarray}
where $F_\nu=\sum_{m,m'}\psi^*_m(\rm{f}_\nu)_{mm'}\psi_{m'}$, $n({\bm \rho}, t)=\sum_m|\psi_m({\bm \rho}, t)|^2$ and $F_{\pm}=F_x\pm i F_y$. To study the modulation-induced dynamics, we solve Eqs.~(\ref{m1})-(\ref{m3}) numerically, starting from a homogeneous density embedded with a small noise in all three components \cite{sai08,wil16}.  


\section{Time modulation of $s$-wave scattering lengths}
\label{tm}
We consider a time dependent $a_j(t)=\bar a_j[1+2\alpha_j\cos (2\omega_j t)]$, where $\bar a_j$ is the mean scattering length, $\alpha_j$ is the modulation amplitude and $2\omega_{j}$ is the driving frequency. Scattering lengths can be periodically modulated by Feshbach resonance \cite{chi10, ngu19,zha20}, or using radio frequency or microwave fields \cite{tsc10, din17, han10,pap10}. We consider three cases: (i) $a_0$ time-dependent and $a_2$ constant, (ii) $a_0$ constant and $a_2$ time-dependent, (iii) both $a_0$ and $a_2$ are time-dependent. These cases can be implemented by independently controlling the two scattering lengths \cite{zha09}. The interaction coefficients $\tilde c_{0, 1}$ for the three cases are
\begin{enumerate}[label=\roman*,align=CenterWithParen]
\item $\tilde c_0(t)=\bar c_0+(2\alpha_{0}\bar g_0/3)\cos(2\omega_0 t)$,\\
 $\tilde c_1(t)=\bar c_1-(2\alpha_{0}\bar g_0/3)\cos(2\omega_0 t)$,\\
 \item $\tilde c_0(t)=\bar c_0+(4\alpha_{2}\bar g_2/3)\cos(2\omega_2 t)$, \\
 $\tilde c_1(t)=\bar c_1+(2\alpha_{2}\bar g_2/3)\cos(2\omega_2 t)$, \\
 and
 \item  $\tilde c_0(t)=\bar c_0+(2\alpha_{0}\bar g_0/3)\cos(2\omega_0 t)+(4\alpha_{2}\bar g_2/3)\cos(2\omega_2 t)$, \\
 $\tilde c_1(t)=\bar c_1-(2\alpha_0\bar g_0/3)\cos(2\omega_0 t)+(2\alpha_2\bar g_2/3)\cos(2\omega_2 t)$,
\end{enumerate}
 where $\bar c_0=(\bar g_0+2\bar g_2)/3$ and $\bar c_1=(\bar g_2-\bar g_0)/3$ with $\bar g_{j}=4\pi\hbar^2\bar a_{j}/(M\sqrt{2\pi}l_z)$. 

The homogeneous solution in the presence of modulation is $\bm \psi(t)=\sqrt{\bar n}{\bm \zeta}\exp(-i\theta(t)/\hbar)$ where $\bm \psi(t)=(\psi_1, \psi_0, \psi_{-1})^T$ and 
\begin{equation}
\theta(t)=\int_{0}^{t}\left[\bar n\tilde c_0(t')+A\tilde c_1(t')\right]dt'+Bt
\end{equation}
where $A=\left[2n_0\left(n_1+n_{-1}\right)+F_z^2+4n_0\sqrt{n_1n_{-1}}\right]/\bar n$ and $B=q(n_1+n_{-1})/\bar n$ with $F_z=n_1-n_{-1}$ and $n_m=\bar n\zeta_m^2$. Now, we introduce 
\begin{equation}
\bm \psi(\bm \rho, t)=\left[\sqrt{\bar n}\bm \zeta+\bm w(t)\cos({\bf k}\cdot{\bm \rho})\right]e^{-i\theta(t)/\hbar},
\end{equation}
in Eqs.~(\ref{m1}-\ref{m3}), where $\bm w(t)=(w_1, w_0, w_{-1})^T$ is the amplitude of modulations and linearize in $\bm w(t )$. Writing $\bm w(t)=\bm u(t)+i\bm v(t)$ where $\bm u=(u_1, u_0, u_{-1})^T$ and $\bm v=(v_1, v_0, v_{-1})^T$ are real-valued vectors, we obtain the two first-order coupled differential equations: 
\begin{eqnarray}
\label{eu1}
    -\hbar\frac{d{\bm v} }{dt}&=&\left(E_k\mathcal I+\mathcal M_1+\mathcal M_2\right){\bm u}\\
    \hbar\frac{d{\bm u} }{dt}&=&\left(E_k\mathcal I+\mathcal M_1+ \mathcal M_3\right){\bm v}
    \label{ev1}
\end{eqnarray}
where $k=|{\bf k}|$, $E_k=\hbar^2k^2/2M$, $\mathcal I$ is a $3\times 3$ identity matrix, $\mathcal M_1$ is a time-independent diagonal matrix depending only on the Zeeman field with elements $(\mathcal M_1)_{11}=(\mathcal M_1)_{33}=q
-B$ and $(\mathcal M_1)_{22}=-B$. Whereas $\mathcal M_{2, 3}$ depends on the initial spinor $\psi_m$, and are real, interaction and time-dependent matrices. Equations (\ref{eu1}) and (\ref{ev1}) can be combined into a Mathieu-like second-order differential equation depending on the system parameters.


\section{Ferromagnetic phase}
\label{fmp}
%
\begin{figure}
\centering
\includegraphics[width= 1.\columnwidth]{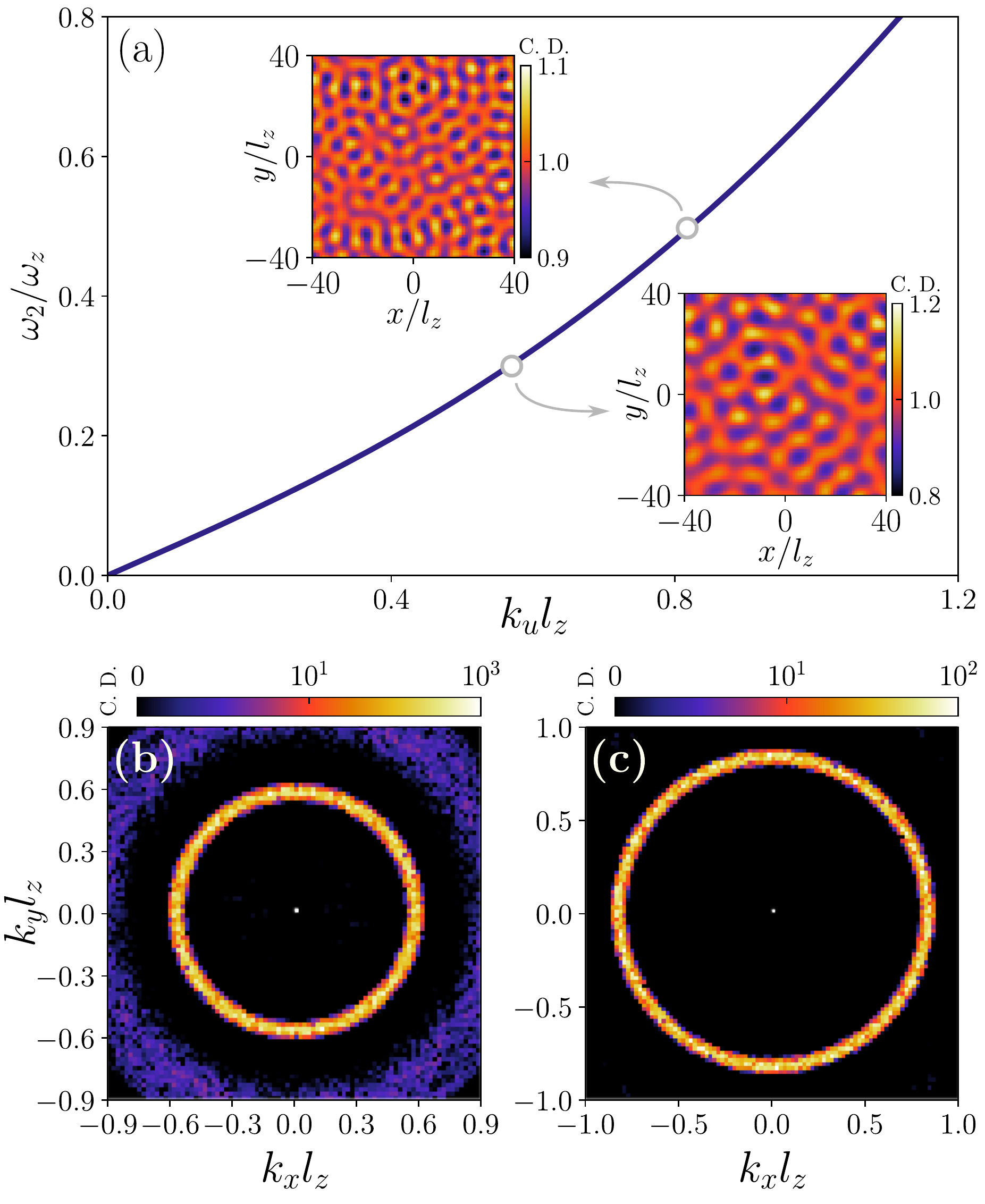}
\caption{\small{(color online). Wavenumber selection of an initial ferromagnetic homogeneous phase for $a_2$ modulation, $\bar c_0\bar n=0.3 \hbar\omega_z$, $\bar c_1\bar n=-0.1 \hbar\omega_z$, $q=-0.3 \hbar\omega_z$, and $\alpha_2=0.4$. (a) The most unstable momentum $k_u$ as a function of driving frequency $\omega_2$. The insets show the numerical results of Faraday patterns, and (b) and (c) show the corresponding condensate momentum density for $\omega_2/\omega_z=0.3$ and $\omega_2/\omega_z=0.5$ at $\omega_z t=330$ and $\omega_z t=250$, respectively. C.D. stands for condensate density, and in (b) and (c), the central peak at $k=0$ is removed for the visibility of the momentum rings.}}
\label{fig:1} 
\end{figure}
For an initial ferromagnetic phase with $\bm \zeta=(1, 0, 0)^{\rm T}$, the matrix $\mathcal M_2$ is diagonal in form with elements $(\mathcal M_2)_{11}=2\bar n[\tilde c_0(t)+\tilde c_1(t)]$, $(\mathcal M_2)_{22}=0$ and $(\mathcal M_2)_{33}=-2\bar n\tilde c_1(t)$, and $\mathcal M_3$ has only one non-zero element $(\mathcal M_3)_{33}=-2\bar n\tilde c_1(t)$. The relevant Mathieu equation is,
\begin{equation}
\label{uf1}
\frac{d^2 u_{1}}{dt^2}+\frac{1}{\hbar^2}\left[\epsilon_{k, 1}^2+4\bar nE_k\alpha_2\bar g_2\cos(2\omega_2 t)\right]u_{1}=0,
\end{equation}
where
\begin{equation}
\label{ef1}
\epsilon_{k, 1}=\sqrt{E_k[E_k+2(\bar c_0+\bar c_1)\bar n]}, 
\end{equation}
is the Bogoliubov dispersion describing the density excitations \cite{sta13}. The dynamical stability of the uniform ferromagnetic phase demands $\bar c_0+\bar c_1>0$ (real $\epsilon_{{\bm k}, 1}$). Since Eq.~(\ref{uf1}) is independent of $\alpha_0$, the ferromagnetic state is immune to the parametric driving of $a_0$. It is because periodic modulation of $a_0$ causes equal and out-of-phase oscillations in $\tilde c_0$ and $\tilde c_1$, which cancel each other.

According to the Floquet theorem, the solutions of Eq.~(\ref{uf1}) are $u_1(t )=b(t) \exp(\sigma t)$, where $b(t ) = b(t + \pi/\omega_2)$ and $\sigma(k,\omega_2, \alpha)$ is called the Floquet exponent. If $\rm{Re}(\sigma) > 0$, the ferromagnetic state is dynamically unstable against the formation of transient density modulations or Faraday patterns [see insets of Fig.~\ref{fig:1}(a)]. The pattern size is determined by the most unstable momentum $k_u$, i.e., momentum for which $\sigma$ is the largest. For vanishing modulation amplitude ($\alpha_2\to 0$), $k_u$ is determined by the resonance $\epsilon_{{\bm k}, 1} = \hbar \omega_2$ and the corresponding Floquet exponent is $\sigma\simeq \bar nE_{k_u}\alpha_2\bar g_2/\hbar^2\omega_2$ \cite{nic07}. Since $\epsilon_{{\bm k}, 1}$ is a monotonously increasing function of $k$, $k_u$ increases with $\omega_2$ (see the unstable momentum rings in Figs.~\ref{fig:1}(b) and \ref{fig:1}(c) for two different frequencies). It implies that the pattern size decreases monotonously with increasing driving frequency $\omega_2$. At longer times, the higher harmonics ($\epsilon_{{\bm k}, 1} = j\hbar \omega_2$ with $j=2, 3, ...$) become relevant, causing the emergence of other rings of higher $|{\bf k}|$ in the momentum density, thus heating and destroying the condensate \cite{bou19,zhan20}.


\section{Polar phase}
\label{pp}

For the polar phase $\bm \zeta_{P}=(0, 1,  0)^{\rm T}$, the chemical potential is $\mu_{2D}=\bar c_0\bar n$ and we have
\begin{eqnarray}
\mathcal M_2=\bar n\begin{bmatrix}
\tilde c_1(t)&0&\tilde c_1(t) \\
0&2\tilde c_0(t)&0 \\
\tilde c_1(t)&0&\tilde c_1(t)
\end{bmatrix}; \ \ \
\mathcal M_3=\bar n\tilde c_1(t)\begin{bmatrix}
1&0&-1 \\
0&0&0 \\
-1&0&1
\end{bmatrix}.\nonumber
\end{eqnarray}
We get the Mathieu-like equations 
\begin{eqnarray}
\label{ep1}
\frac{d^2 u_0 }{d t^2}+\frac{1}{\hbar^2}E_k\left[E_k+2\bar n\tilde c_0(t)\right] u_0=0\\
\frac{d^2 u_+ }{d t^2}+\frac{1}{\hbar^2}\left(E_k+q\right)\left[E_k+q+2\bar n\tilde c_1(t)\right] u_+=0,
\label{ep2}
\end{eqnarray}
where $u_+=u_1+u_{-1}$. Unlike the ferromagnetic case, both the density and spin modes,
\begin{eqnarray}
\epsilon_{k, 0}&=&\sqrt{E_k(E_k+2\bar c_0\bar n)} \\
\epsilon_{k, \pm 1}&=&\sqrt{(E_k+q)(E_k+q+2\bar c_1\bar n)},
\label{spm}
\end{eqnarray}
become important for the dynamics. $\epsilon_{k, 0}$ corresponds to the density modulations (phonons) and the degenerate $\epsilon_{{\bf k}, \pm 1}$ modes associated with the elementary process of $({\bm 0}, 0)+({\bm 0}, 0)\leftrightarrow({\bf k}, \pm 1)+(-{\bf k}, \mp 1)$. When $q=0$ and $\bar c_0=\bar c_1$, all three modes are degenerate. The dynamical stability of the homogeneous polar phase demands $\bar c_0>0$ and $q(q+2\bar c_1\bar n)\geq 0$ and $q+\bar c_1\bar n\geq 0$. Below we consider the three cases of modulation. In the numerical calculations of Eqs.~(\ref{m1})-(\ref{m3}), the initial polar phase is embedded with a noise field $\delta({\bm \rho})$ populating the vacuum modes in the limit $q\to\infty$, based on the  truncated Wigner prescription \cite{bar11, wil16},
\begin{eqnarray}
		\delta({\bm \rho})=\frac{1}{\sqrt{V}}\sum_{\bf k}
		\begin{pmatrix} 
			\alpha_{\bf k}^{+1}\exp(i{\bf k}\cdot{\bm\rho}) \\ \alpha_{\bf k}^{0}u_{\bf k}\exp(i{\bf k}\cdot{\bm\rho})+\alpha_{\bf k}^{0*}v_{\bf k}\exp(-i{\bf k}\cdot{\bm\rho}) \\ \alpha_{\bf k}^{-1}\exp(i{\bf k}\cdot{\bm\rho})
		\end{pmatrix},
	\end{eqnarray}
where $ V $ is the volume, $ \alpha_{{\bf k}}^{m} $ are complex Gaussian random variables with zero mean and satisfies $ \left\langle\alpha_{\bf k}^{m*}\alpha_{\bf k'}^{m'}\right\rangle = (1/2)\delta_{mm'}\delta_{\bf k\bf k'}$ and the amplitudes are given by,
	\begin{eqnarray}
		u_{\bf k} = \sqrt{\frac{E_k+c_0\bar n}{2\sqrt{E_k(E_k+2c_0\bar n)}}}-\frac{1}{2} \\
		v_{\bf k} = \sqrt{1-u_{\bf k}^2}.
	\end{eqnarray} 
We also found qualitatively similar results for an initial state 
\begin{eqnarray}
		\bm \psi(\bm \rho,t=0)=
	\begin{pmatrix} 
		p_{+1}\exp(i\theta_1) \\
		[1-p_{+1}^2-p_{-1}^2]\exp(i\theta_{0}) \\
		p_{-1}\exp(i\theta_{-1})
	\end{pmatrix}
	\end{eqnarray}
with a noise from a uniform distribution, where $p_{\alpha} $ and $ \theta_{\alpha} $ are random numbers with $p_{\pm 1}\ll 1$.  
 
\subsection{$a_0$ Modulation}
%
%
\begin{figure}
\centering
\includegraphics[width= 1.\columnwidth]{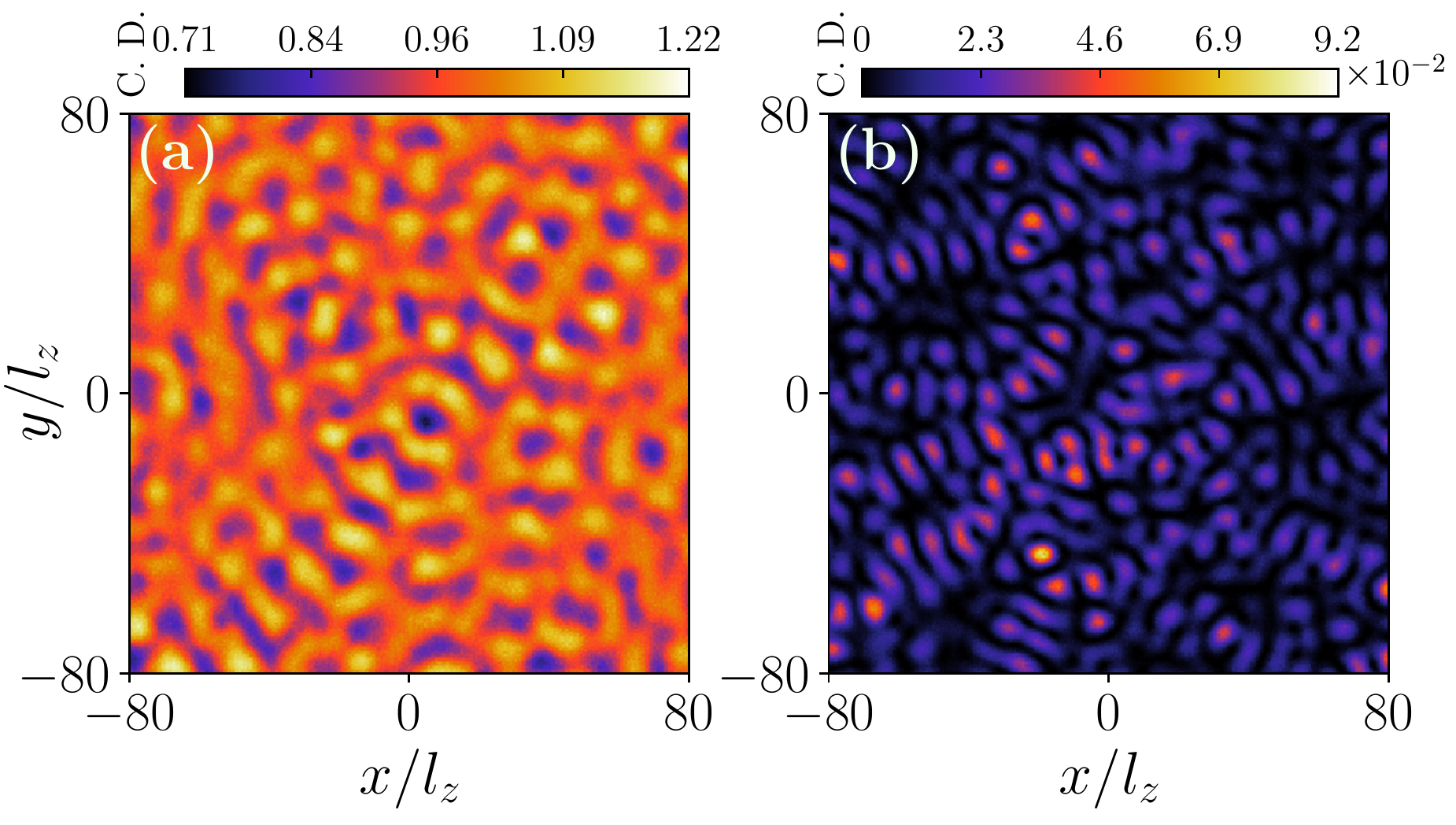}
\caption{\small{Density patterns for $a_0$ modulation on an initial polar phase. The parameters are $q=0$, $\bar c_0\bar n =\bar c_1\bar n=0.2\hbar \omega_z$, $\alpha_0 = 0.4$, and $\omega_0/\omega_z = 0.2$ at $\omega_z t=600$. The density pattern in (b) $m=\pm 1$ differs from that of (a) $m=0$. C.D. stands for condensate density.}}
\label{fig:2} 
\end{figure}
\begin{figure}
\centering
\includegraphics[width= 1.\columnwidth]{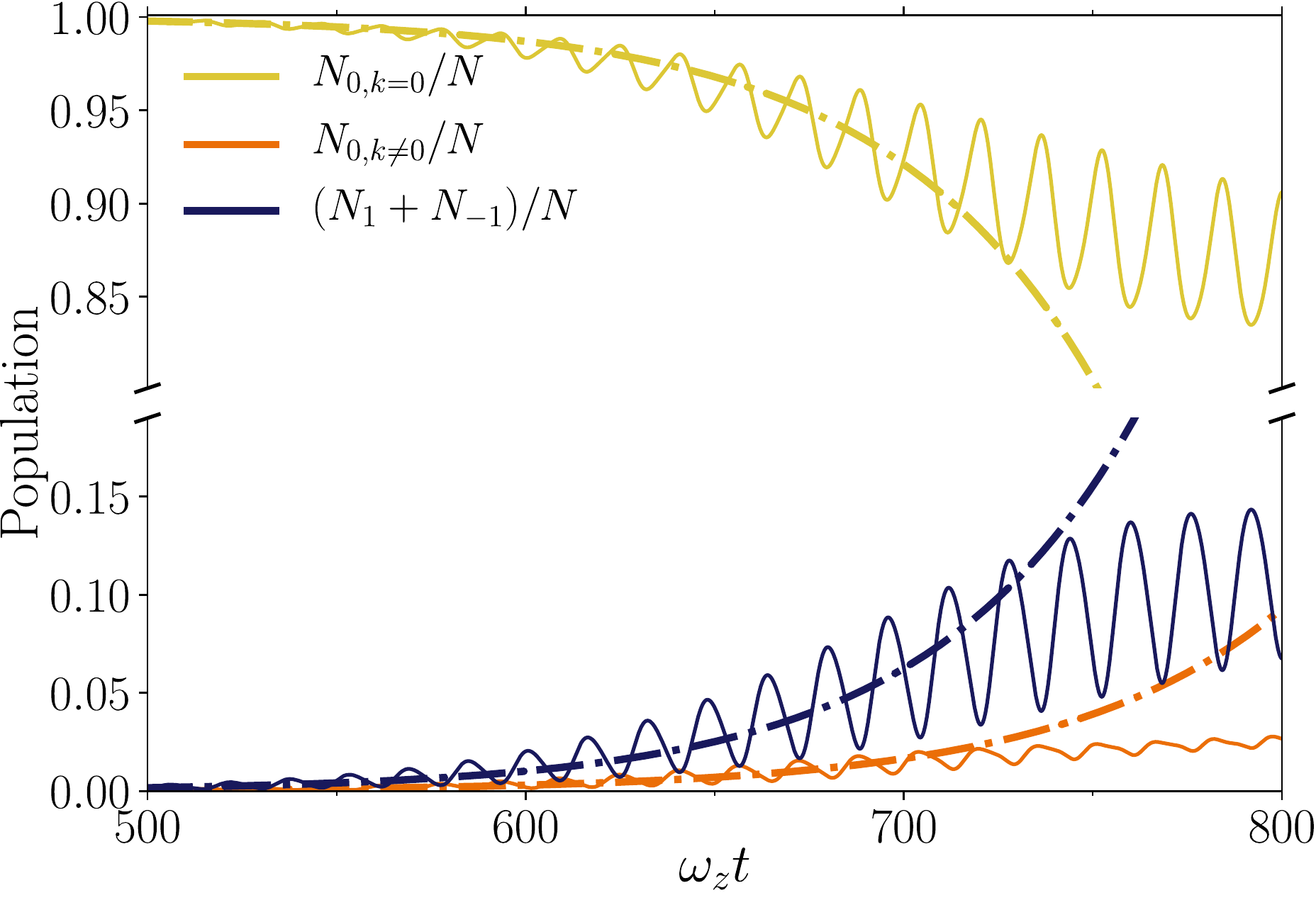}
\caption{\small{Population dynamics for $a_0$ modulation and $q=0$. Other parameters are $\bar c_0\bar n =\bar c_1\bar n=0.2\hbar \omega_z$, $\alpha_0 = 0.4$, and $\omega_0/\omega_z = 0.2$. The population in each component is $N_m(t) = \int |\psi_m(x,y,t)|^2 dxdy$, the total population $N=\sum_mN_m$, the population with $k\neq 0$ in $m=0$ is $N_{0,k\neq 0} = \int_{k\neq 0} dk_xdk_y|\tilde \psi_{0}(k_x, k_y,t)|^2$ and the zero-momentum population in $m=0$ is $N_{0,k=0}$. The dashed-dotted lines show the exponential fit determined by the Floquet exponent, and the solid lines are from the numerical calculations of NLGPEs.}}
\label{fig:3} 
\end{figure}
%
For $a_0$ modulation, the Mathieu Eqs.~(\ref{ep1}) and (\ref{ep2}) become
\begin{align}
\label{epn1}
\frac{d^2 u_0 }{d t^2}+\frac{1}{\hbar^2}\left[\epsilon_{{k}, 0}^2+\frac{4\bar nE_k\alpha_{0}\bar g_0}{3}\cos(2\omega_0 t)\right] u_0=0\\
\frac{d^2 u_+ }{d t^2}+\frac{1}{\hbar^2}\left[\epsilon_{{k}, \pm 1}^2-\frac{4\bar n(E_k+q)\alpha_{0}\bar g_0}{3}\cos(2\omega_0 t)\right] u_+=0.
 \label{epn2}
\end{align}
For a given $\omega_0$ and in the limit $\alpha_0\to 0$, the resonances $\epsilon_{k, 0} = \hbar \omega_0$ and $\epsilon_{k, \pm 1} = \hbar \omega_0$ provide us two unstable momenta $k_u^{(0)}$ and $k_u^{(+)}$ with Floquet exponents $\sigma^{(0)}\simeq \bar nE_{k_u^{(0)}}\alpha_0\bar g_0/3\hbar^2\omega_0$ and $\sigma^{(+)}\simeq \bar n(E_{k_u^{(+)}}+q)\alpha_0\bar g_0/3\hbar^2\omega_0$. Using $k_u^{(0)}$ and $k_u^{(+)}$ obtained from the resonance conditions, we rewrite the Floquet exponents as
 \begin{eqnarray}
 \label{s0}
\sigma^{(0)}=\frac{\alpha_0\bar n\bar g_0}{3\hbar^2\omega_0}\left[\sqrt{(\bar n\bar c_0)^2+\hbar^2\omega_0^2}-\bar n\bar c_0\right],\\
 \sigma^{(+)}=\frac{\alpha_0\bar n\bar g_0}{3\hbar^2\omega_0}\left[\sqrt{(\bar n\bar c_1)^2+\hbar^2\omega_0^2}-\bar n\bar c_1\right].
 \label{s1}
 \end{eqnarray}
When $\bar c_0=\bar c_1$, $\sigma^{(0)}=\sigma^{(+)}$ and interestingly, $\sigma^{(+)}$ is independent of the quadratic Zeeman field $q$, which has interesting consequences as we discuss later. 

\begin{figure}
\centering
\includegraphics[width= 1.\columnwidth]{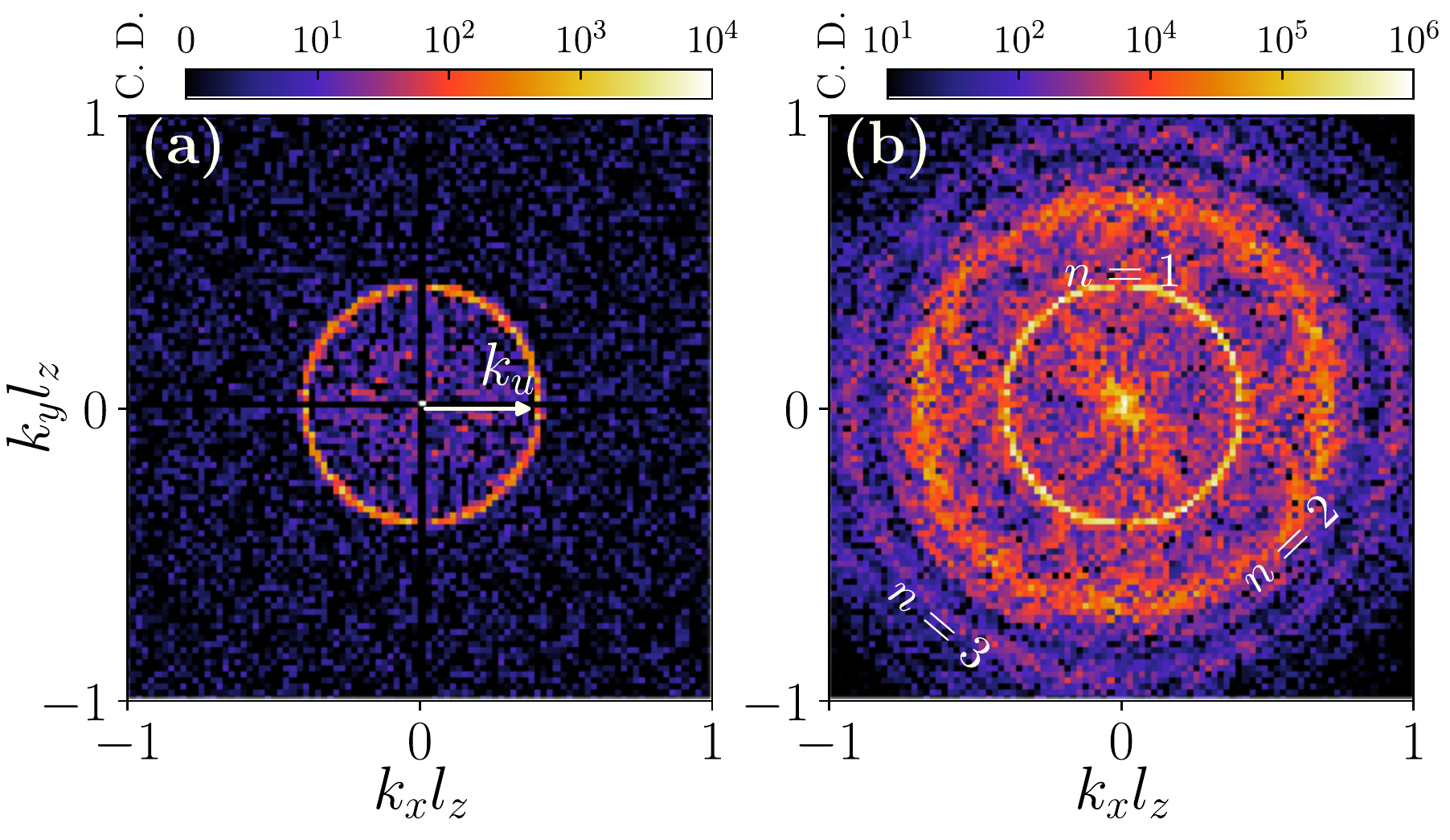}
\caption{\small{The momentum density of $m=0$ component at (a) $\omega_zt=300$ and (b) $\omega_zt=650$ for $\bar c_0\bar n =\bar c_1\bar n=0.2\hbar \omega_z$, $\alpha_0 = 0.4$, and $\omega_0/\omega_z = 0.2$. The momentum ring in (a) and the primary one in (b) is given by the resonance $\epsilon_{{\bm k}, 0} = \hbar \omega_0$. Two higher harmonics can also be seen in (b). C.D. stands for the condensate density, and the central peak at $k=0$ is removed for the visibility of the momentum rings.}}
\label{fig:3a} 
\end{figure}

\subsubsection{Degenerate modes}
\label{dm}
%
\begin{figure}
\centering
\includegraphics[width= .92\columnwidth]{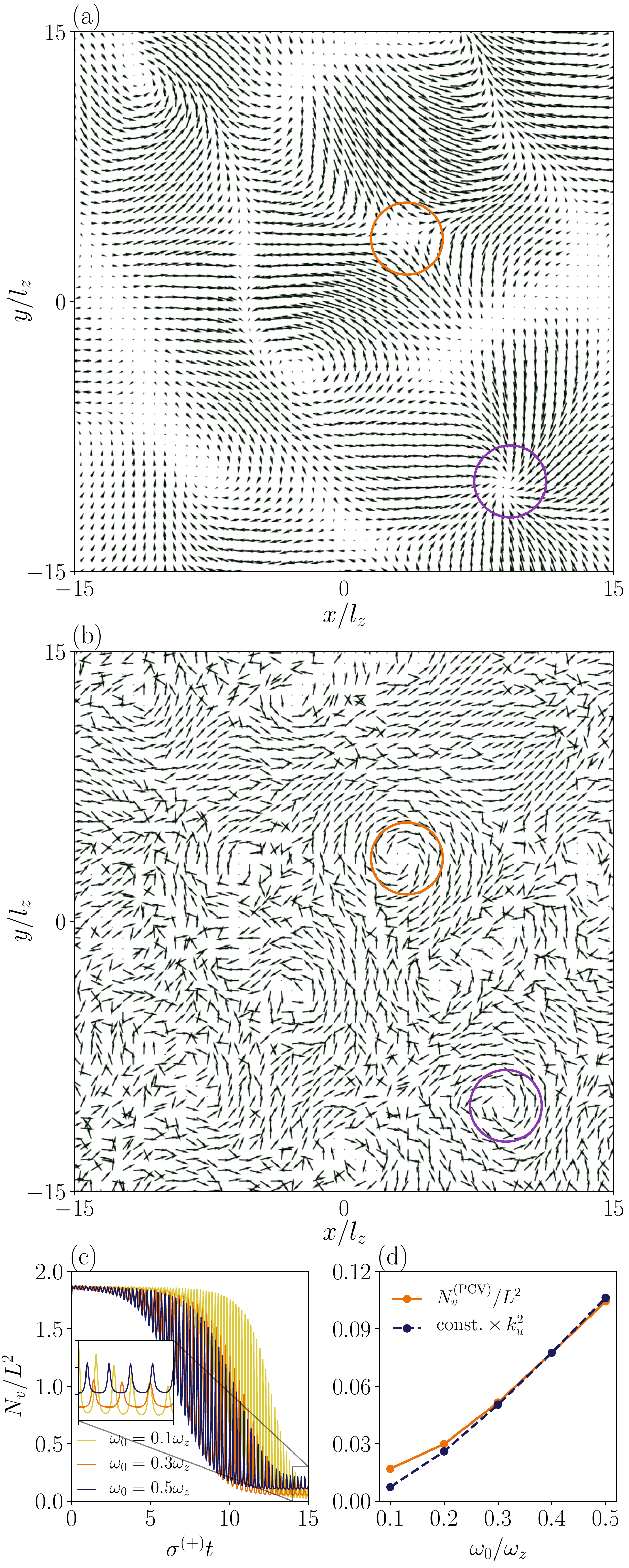}
\caption{\small{Spin textures for $a_0$ modulation and $q=0$. (a) The transverse spin density ${\bm F}_\perp(x, y)$ and (b) the velocity field of ${\bm F}_\perp(x, y)=(F_x, F_y)$, i.e., $\vec\nabla\phi/|\vec\nabla\phi|$, where $\phi=\arctan(F_x/F_y)$ at $\omega_z t=670$ or $\sigma^{(+)} t = 14.74$. Other parameters are the same as in Fig.~\ref{fig:3}. Both vortices and antivortices are marked in (b). (c) The dynamics of the vortex density $N_v/L^2$, where $L$ is the numerical box size, is obtained by averaging over ten realizations of noises. It decreases in time and eventually saturates. The saturated vortex density (before the condensate is destroyed, taken around $\sigma^{(+)}t=15$) as a function of driving frequency $\omega_0$ is shown in (d), which exhibits a quadratic dependence on the most unstable momentum. The latter is expected, considering the Q2D nature of the condensate.}}
\label{fig:4} 
\end{figure}
%
When all three modes are degenerate (when $q=0$ and $\bar c_1=\bar c_0$), $k_u^{(0)}=k_u^{(+)}=k_u$ and $\sigma^{(0)}=\sigma^{(+)}$. It implies that both density and spin modes get populated simultaneously by periodic driving. The unstable density mode ($\epsilon_{{\bm k}, 0}$) leads to the formation of Faraday patterns [see Fig.~\ref{fig:2}(a)] or equivalently the exponential growth of atoms with finite momenta in $m=0$ component, shown as $N_{0, k\neq 0}$ in Fig.~\ref{fig:3}. The unstable $\epsilon_{{\bm k}, \pm 1}$ modes initiate the population transfer from $m=0$ to $m=\pm 1$, leading to the spin dynamics seen in Fig.~\ref{fig:3}. At the initial stage, the population in $m=\pm 1$ grows exponentially at a rate determined by $\sigma^{(+)}$ and exhibits oscillations of frequency $2\omega_0$. Eventually, a density pattern also develops in $m=\pm1$ components [see Fig.~\ref{fig:2}(b)]. 

The momentum density of $m=0$ is shown in Fig.~\ref{fig:3a} at two different instances. At the early stages of the pattern formation, the most unstable momenta are given by the resonance condition $\epsilon_{k, 0} =\hbar \omega_0$, which we marked by $k_u$ in Fig.\ref{fig:3a}(a). Although the next leading unstable momenta come from the second harmonics $\epsilon_{k, 0} =2 \hbar \omega_0$, there are also contributions from processes in which a pair of atoms, each from $m=+1$ and $m=-1$ with momenta of magnitudes $(k_u, -k_u)$ or $(\pm k_u, \pm k_u)$ scatter into $(+{\bf k}, -{\bf k})$ or $({\bf k}, {\bf k'})$ with $|{\bf k}+{\bf k'}|=2k_u$. All such ${\bf k}$ and ${\bf k'}$ lead to a disk-shaped pattern in the momentum space of the condensate wavefunction in addition to the harmonics, but with lesser amplitudes, as shown in Fig.~\ref{fig:3a}(b). At longer times, the condensate gets destroyed by heating.

%
{\em Magnetization dynamics}: The initial polar condensate has a null spin density vector ${\bm F}$. The spin-mixing dynamics in Fig.~\ref{fig:3} leads to the emergence of spin textures shown in Fig.~\ref{fig:4}(a). Strikingly, the velocity field of transverse spin density vector ${\bm F}_\perp(x, y)$ in Fig.~\ref{fig:4}(b) reveals the formation of polar-core vortices (PCVs) and anti-vortices \cite{iso01, miz04, sad06}. They are marked by circles in Figs.~\ref{fig:4}(a) and \ref{fig:4}(b). The core of a PCV is filled with $m=0$ atoms with no vorticity, and the surrounding $m=+1$ and $m=-1$ atoms have opposite vorticity. Figure~\ref{fig:4}(c) shows the time dependence of vortex density $N_v/L^2$, where $L^2$ is the area of the numerical box we use. The vortex number $N_v$ is determined by computing the phase winding at the smallest loops defined by the grid size. At $t=0$, the initial random noise in $m=\pm 1$ components contributes to the vortex number $N_v$, which eventually decays over time as the PCVs materialize. $N_v$ decreases until the number of PCVs reaches a steady value $N_v^{\rm{(PCV)}}$ and the decay rate is determined by $\sigma^{(+)}$.

Since the amplitude of spin texture oscillates in time with the driving frequency, $N_v$ exhibits the same. When the spin-texture amplitude is tiny, the $N_v$ becomes large from the noise contribution. If the amplitude is sufficiently large, the noise is overshadowed. Once $N_v$  reaches a steady value, the minima shown in the inset of Fig.~\ref{fig:4}(c)] approximately provide us with the number of PCVs. Interestingly, $N_v^{\rm{(PCV)}}$ is determined by the unstable momentum $k_u$ as shown in Fig.~\ref{fig:4}(c)] and as expected it exhibits $k_u^2$ behaviour. It implies that the larger the modulation frequency, the denser the spin-vortex gas. At longer times ($\sigma^{(+)}t>16$), we observed that $m=0$ homogeneous condensate gets significantly depleted, and PCVs disintegrated into independent gases of vortices in $m=1$ and $m=-1$ components.
 
\begin{figure}
\centering
\includegraphics[width= .95\columnwidth]{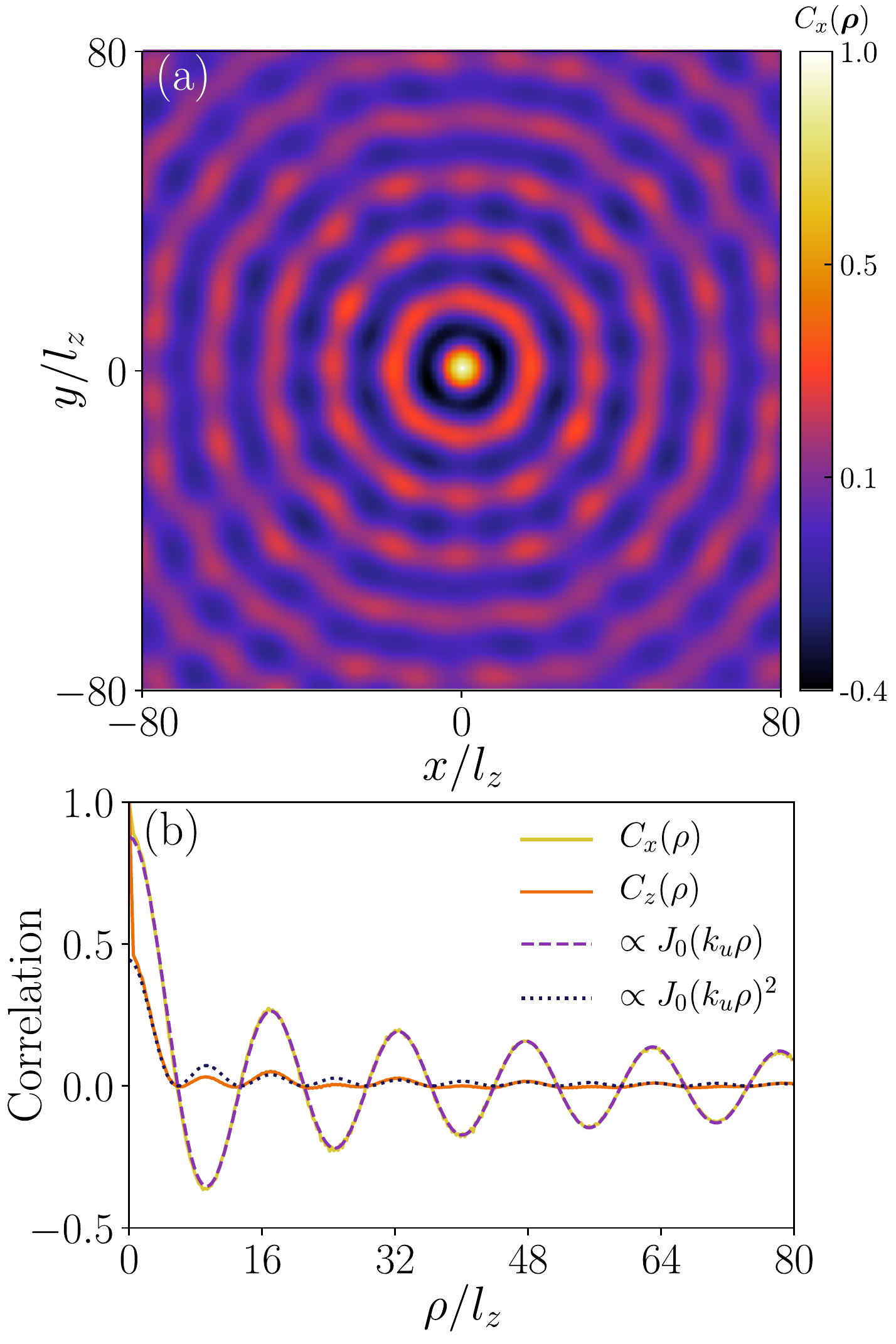}
\caption{\small{Spin-spin correlations for $a_0$ modulation and $q=0$. (a) The transverse magnetization correlation $C_{x}({\bm \rho}, t)$ in the $xy$-plane and (b) the radial correlation function $C_{x, z}(\rho)$ at $\omega_zt=500$. Other parameters are the same as in Fig.~\ref{fig:3}. We see that $C_{x,y}(\rho)\propto J_0(k_u\rho)$ and $C_{z}(\rho)\propto J_0^2(k_u\rho)$. The plots are obtained by taking an average of results from ten different realizations of initial noise.}}
\label{fig:5} 
\end{figure}

 Further, we analyze the scaled spin-spin correlations,
\begin{equation}
C_{\alpha}({\bm \rho}, t)=\frac{1}{N_{\alpha}}\iint d{\bm \rho}'F_{\alpha}({\bm \rho}+{\bm \rho}', t)F_{\alpha}({\bm \rho}', t),
\label{cor1}
\end{equation}
where $\alpha\in\{x, y, z\}$ and $N_{\alpha}=\iint d{\bm \rho}F_{\alpha}({\bm \rho})^2$. $C_{x}({\bm \rho}, t)$ at an instant well before the condensate is destroyed is shown in Fig.~\ref{fig:5}(a). The radial correlations along the transverse and longitudinal magnetization densities $C_{\alpha}(\rho)=(1/2\pi)\int_0^{2\pi}C_{\alpha}({\bm \rho})d\theta$ are found to be governed by Bessel function $J_0(k_u\rho)$, where $\rho=|{\bm \rho}|$ and in particular, $C_{x,y}(\rho)\propto J_0(k_u\rho)$ and $C_{z}(\rho)\propto J_0^2(k_u\rho)$ [see Fig.~\ref{fig:5}(b)]. Similar Bessel correlations are predicted in spinor condensates subjected to quantum quenches \cite{lam07, sai07a,bar11}.

%

\subsubsection{Non-degenerate modes}
\label{ndm}
\begin{figure}
\centering
\includegraphics[width= 0.96\columnwidth]{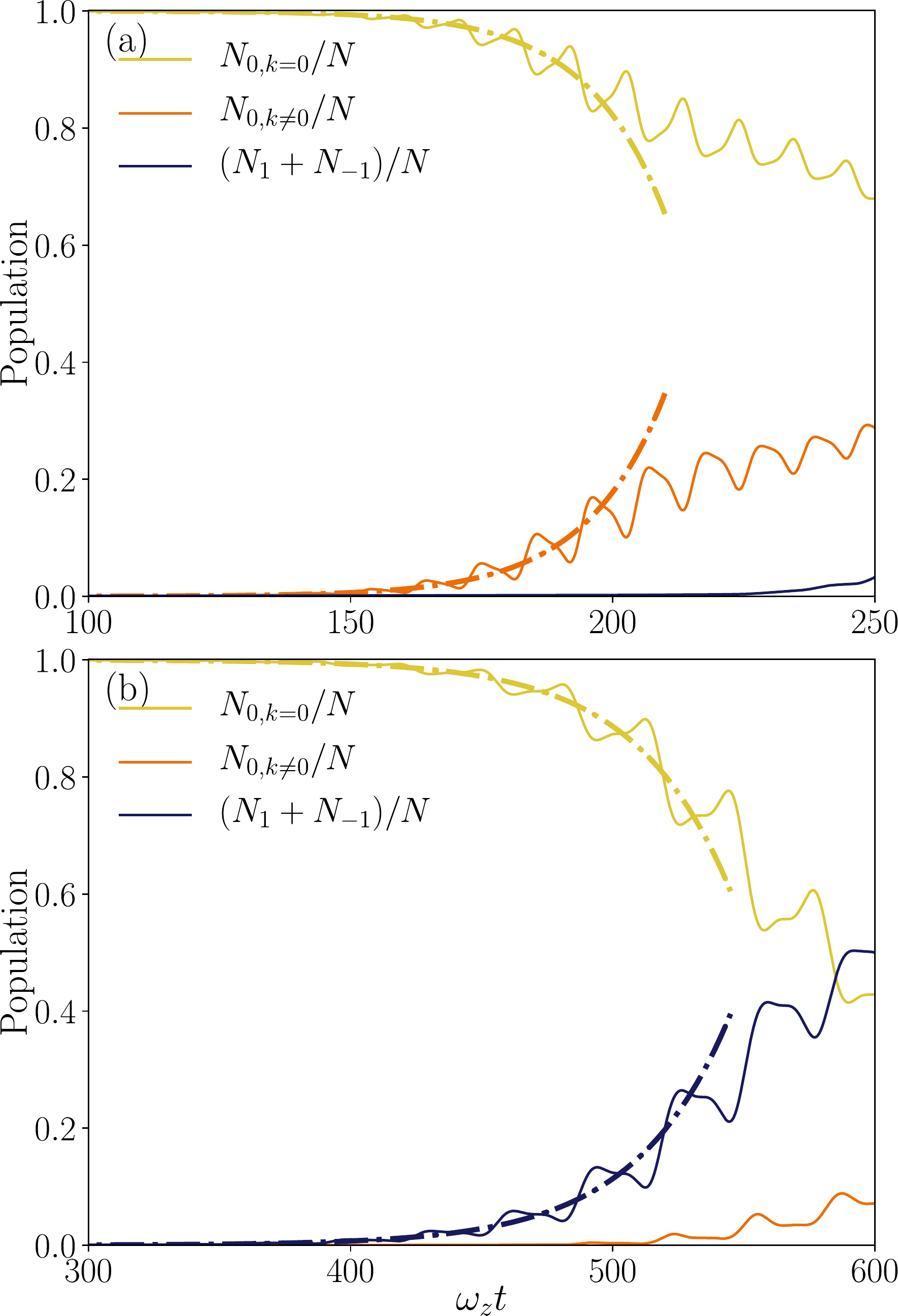}
\caption{\small{The population dynamics for $a_0$ modulation and $q=0$. (a) $\bar c_0\bar n=0.2\hbar \omega_z$, $\bar c_1\bar n=4\hbar \omega_z$, $\alpha_0=0.03$ and $\omega_0/\omega_z=0.3$. (b) $\bar c_0\bar n=0.4\hbar \omega_z$, $\bar c_1\bar n=0.05\hbar \omega_z$, $\alpha_0=0.25$ and $\omega_0/\omega_z=0.1$. In (a) $N_{0,k\neq 0}$ outgrows $N_{\pm 1}$ and vice versa in (b). The solid lines are the numerical results, and dotted-dashed lines are the exponential fit provided by $\sigma^{(0)}$.}}
\label{fig:6} 
\end{figure}

\begin{figure}
\centering
\includegraphics[width= .96\columnwidth]{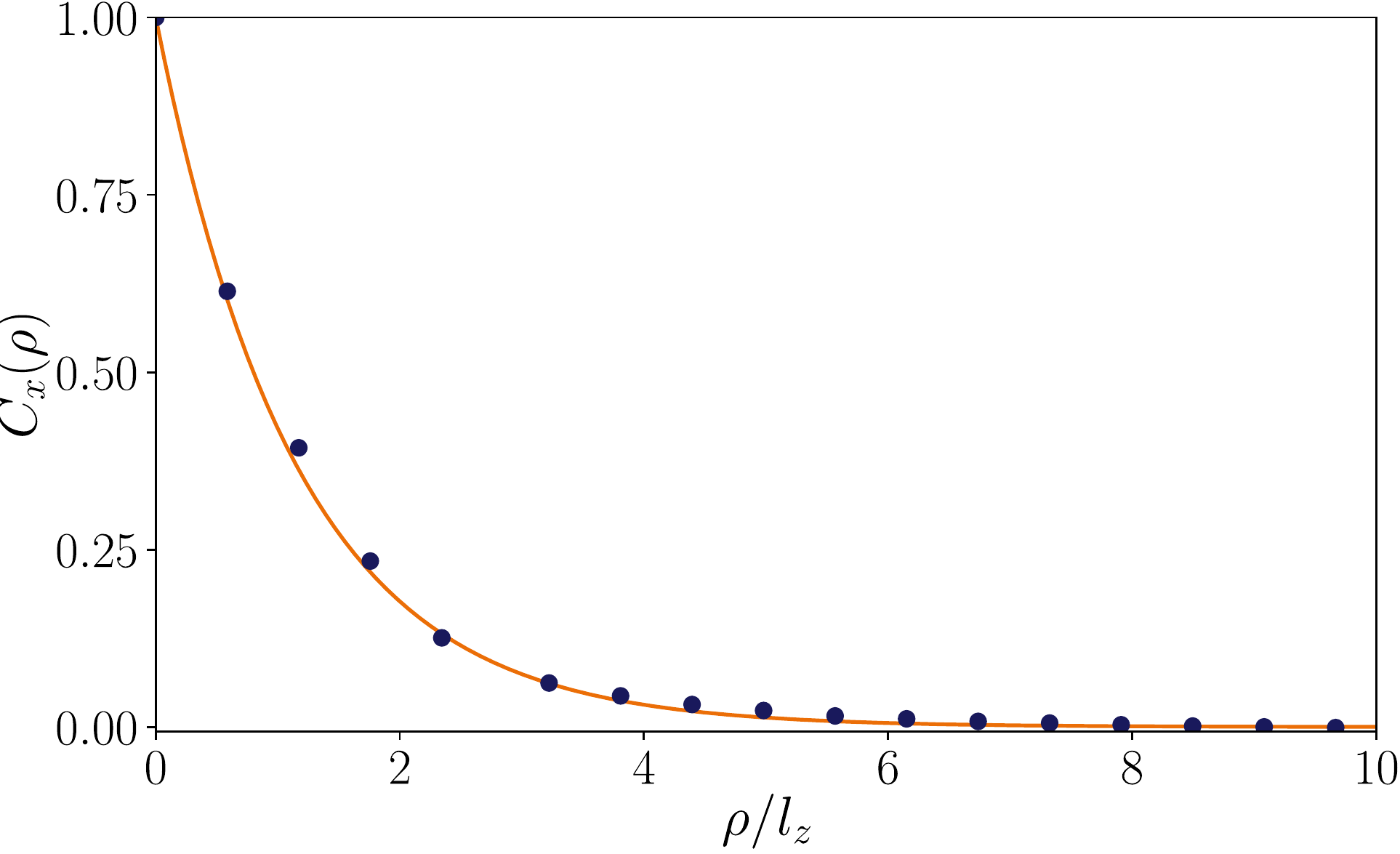}
\caption{\small{The radial transverese spin-spin correlation function $C_{x}(\rho)$ for $\bar c_0\bar n=0.2\hbar \omega_z$, $\bar c_1\bar n=4\hbar \omega_z$, $\alpha_0=0.03$ and $\omega_0/\omega_z=0.3$, exhibiting exponential decay at an instant $\omega_zt=240$. The dots are obtained from average of results from ten different realizations of initial noise. The solid line is an exponential fit. The corresponding population dynamics is shown in Fig.~\ref{fig:6}(a).}}
\label{fig:6a} 
\end{figure}

For $q\neq 0$ or $\bar c_1\neq\bar c_0$, the degeneracy between the density ($\epsilon_{k, 0}$) and the spin ($\epsilon_{k, \pm 1}$) modes is lifted and consequently, $k_u^{(0)}\neq k_u^{(+)}$. First, we consider $q=0$ and $\bar c_1\neq\bar c_0$. If $\sigma^{(0)}$ and $\sigma^{(+)}$ are comparable, the dynamics remains qualitatively the same as in the case of degenerate modes. So we consider the two extreme scenarios: $\bar c_0\ll\bar c_1$ and $\bar c_1\ll\bar c_0$. For $\bar c_0\ll\bar c_1$, the density mode is the soft one and consequently $\sigma^{(0)}\gg\sigma^{(+)}$. In that case, the Faraday pattern in $m=0$ forms well before the spin-mixing dynamics takes place [see the dynamics of $N_{0,k\neq 0}$ and $N_{\pm 1}$ in Fig.~\ref{fig:6}(a)]. The initial growth of $N_{0,k\neq 0}$ is determined by $\sigma^{(0)}$ [dotted-dashed line in Fig.~\ref{fig:6}(a)]. By the time the population transfer to $m=\pm 1$ takes place, the homogeneous condensate in $m=0$ is significantly altered and depleted, making our linear stability analysis invalid. In contrast, for $\bar c_0\gg\bar c_1$, $N_{\pm 1}$ out-grows $N_{0,k\neq 0}$ [see Fig.~\ref{fig:6}(b)] and the initial dynamics is governed by $\sigma^{(+)}$. The spin-dynamics leads to spin textures and the formation of PCVs. The spin-spin correlations have different behavior for the two extreme cases. For $\bar c_0\ll\bar c_1$, $C_{x, y, z}(\rho)$ decays exponentially [see Fig.~\ref{fig:6a}] with a correlation length of the order of a spin healing length, which is proportional to $\propto 1/\sqrt{\bar c_1\bar n}$, whereas, for $\bar c_1\ll\bar c_0$, they exhibit a Bessel function dependence of $C_{x,y}(\rho)=J_0(k_u^{(+)}\rho)$ and $C_{z}(\rho)=J_0^2(k_u^{(+)}\rho)$. 
%
 
\begin{figure}
\centering
\includegraphics[width= 1.\columnwidth]{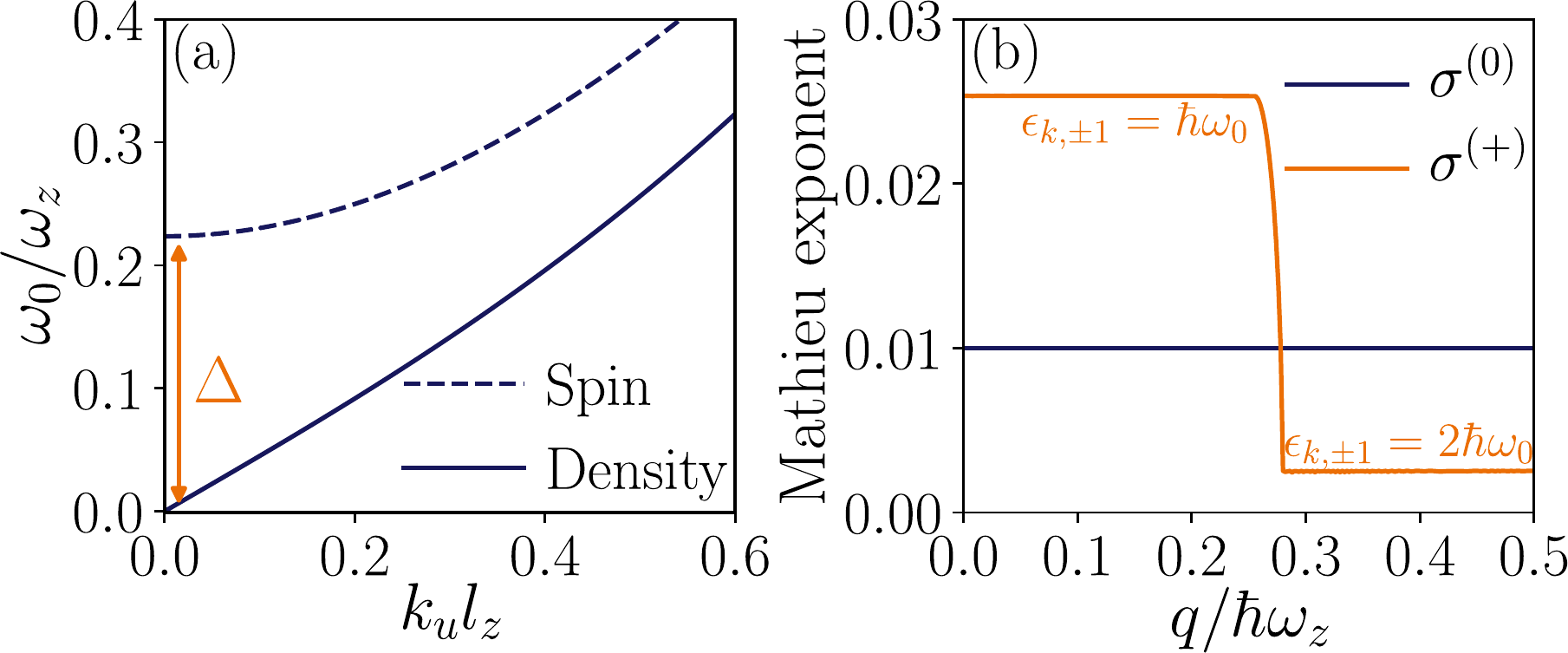}
\caption{\small{(a) The Bogoliubov spectrum for $\bar c_0\bar n=\bar c_1\bar n=0.2\hbar \omega_z$, $\omega_0/\omega_z=0.3$, and $q = 0.1\hbar\omega_z$. The unstable momenta as a function of $\omega_0$ map the spectrum in the limit $\alpha\to 0$. (b) The Mathieu exponents of density and spin modes as a function of $q$ for $\bar c_0\bar n=0.4 \hbar\omega_z$, $\bar c_1\bar n=0.05 \hbar\omega_z$, $\omega_0/\omega_z=0.3$ and $\alpha_0=0.3$. $\Delta$ is the gap of the spin modes at $k=0$. In (b), the primary resonance associated with the spin mode changes from $\epsilon_{{\bm k}, \pm 1}=\hbar\omega_0$ to $\epsilon_{{\bm k}, \pm 1}=2\hbar\omega_0$ as a function of $q$, leading to a significant decrease in the Mathieu exponent $\sigma^{(+)}$.}}
\label{fig:7} 
\end{figure}
{\em Non-zero $q$}. The nature of dynamics also depends critically on the quadratic Zeeman field $q$. A finite $q$ not only lifts the degeneracy of the modes but also introduces a gap, $\Delta=\sqrt{q(q+2\bar c_1\bar n)}$ in the spin modes [see Fig.~\ref{fig:7}(a)]. It means that there is no spin-mixing dynamics if the driving frequency lies below the gap ($\omega_0<\Delta$), and the periodic modulation only leads to forming the Faraday pattern in $m=0$. In contrast, for $\omega_0\geq\Delta$, both spin and density modes contribute to the dynamics. Since the Mathieu exponents are independent of $q$ [see Eqs.~(\ref{s0}) and (\ref{s1})], both modes are unstable simultaneously for $\bar c_1=\bar c_0$. In that case, the dynamics is identical to the case of degenerate modes, but the momenta governing the Faraday pattern and the vortex density are different, i.e., $k_u^{(0)}\neq k_u^{(+)}$. The spatial dependence of spin-spin correlations is governed by $k_u^{(+)}$ via the Bessel function.

Even though the Mathieu exponents are independent of $q$, the above results indicate an implicit dependence of $q$ on the dynamics for a fixed $\omega_0$. For instance, for sufficiently small $q$ such that $\Delta<\omega_0$, both modes are unstable simultaneously, whereas for sufficiently large $q$ such that $\Delta>\omega_0$, only the density mode is unstable. A similar scenario, as shown in Fig.~\ref{fig:7}(b), also emerges for $\bar c_1\ll\bar c_0$. For small $q$, the spin dynamics dominates ($\sigma^{(+)}\gg\sigma^{(0)}$) and for large $q$, only the Faraday pattern is formed. When the spin-mixing dynamics dominates ($\bar c_1\ll\bar c_0$), the spin-spin correlations are again governed by the Bessel functions as before.
 
\subsection{$a_2$ Modulation}
%
%
\begin{figure}
\centering
\includegraphics[width= .95\columnwidth]{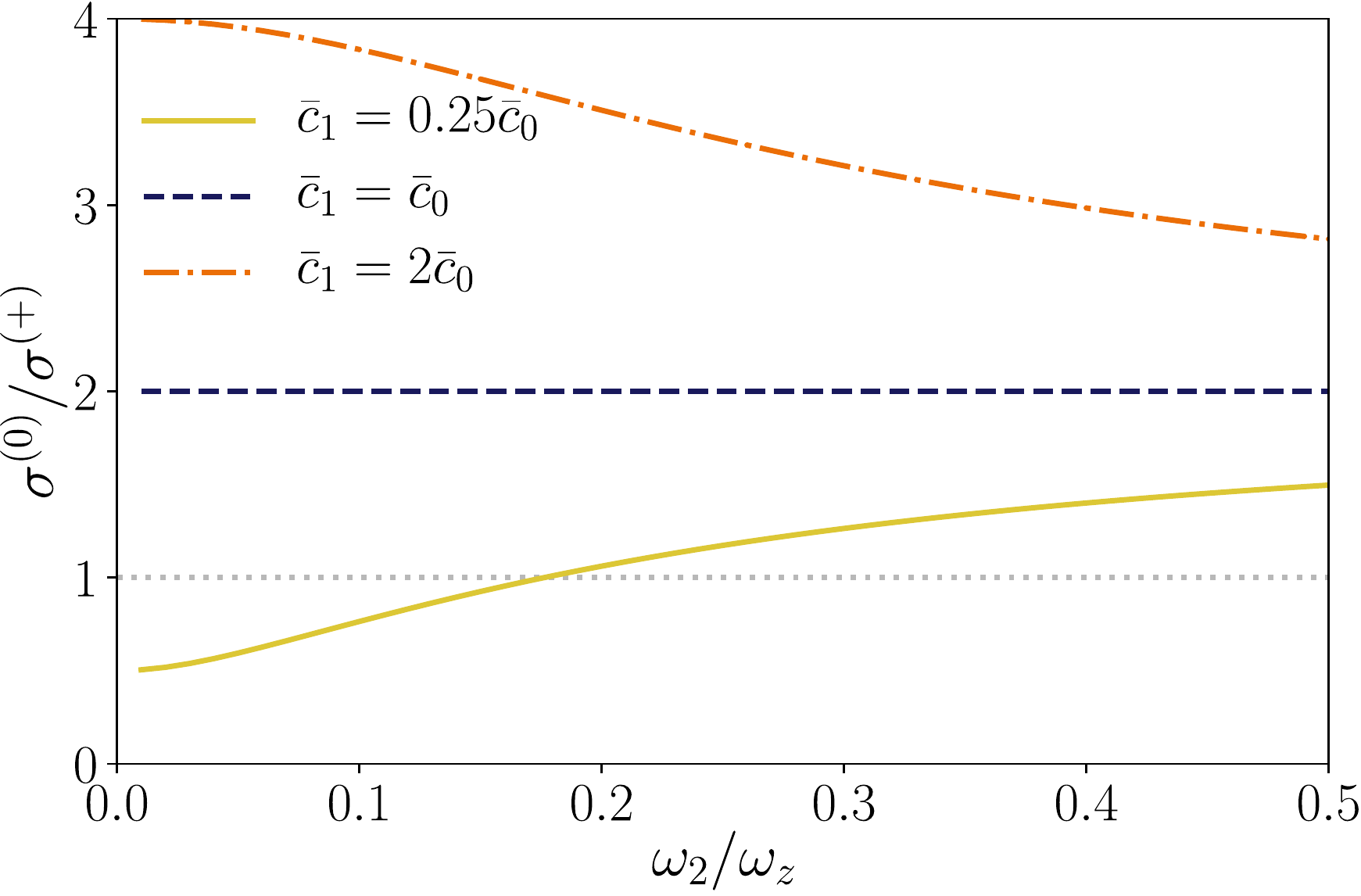}
\caption{\small{The results for $a_2$ modulation. $\sigma^{(0)}/\sigma^{(+)}$ as a function of $\omega_2$ for $\bar c_0\bar n=0.2\hbar\omega_z$ and $q=0$ with different $\bar c_1/\bar c_0$. For $\bar c_0=\bar c_1$, we get $\sigma^{(0)}/\sigma^{(+)}=2$ (dashed line), and for $\bar c_1<\bar c_0$ (solid line), the value of $\sigma^{(0)}/\sigma^{(+)}$ crosses one, which indicates there is a change in the behavior of dynamics above and below a critical driving frequency $(\omega_2\approx 0.18\omega_z)$.}}
\label{fig:8} 
\end{figure}
 For $a_2$ modulation, the Mathieu Eqs.~(\ref{ep1}) and (\ref{ep2}) become
\begin{align}
\label{epna1}
\frac{\mathrm{d}^2 u_0 }{\mathrm{d} t^2}+\frac{1}{\hbar^2}\left[\epsilon_{{k}, 0}^2+\frac{8\bar nE_k\alpha_{2}\bar g_2}{3}\cos(2\omega_2 t)\right] u_0=0\\
 \frac{\mathrm{d}^2 u_+ }{\mathrm{d} t^2}+\frac{1}{\hbar^2}\left[\epsilon_{{k}, \pm 1}^2+\frac{4\bar n(E_k+q)\alpha_{2}\bar g_2}{3}\cos(2\omega_2 t)\right] u_+=0.
 \label{epna2}
\end{align}
The Mathieu exponents for the density and spin modes are obtained as, 
 \begin{eqnarray}
 \label{sa0}
\sigma^{(0)}=\frac{2\alpha_2\bar n\bar g_2}{3\hbar^2\omega_2}\left[\sqrt{(\bar n\bar c_0)^2+\hbar^2\omega_2^2}-\bar n\bar c_0\right],\\
 \sigma^{(+)}=\frac{\alpha_2\bar n\bar g_2}{3\hbar^2\omega_2}\left[\sqrt{(\bar n\bar c_1)^2+\hbar^2\omega_2^2}-\bar n\bar c_1\right].
\label{sa1}
\end{eqnarray}
Comparing to the case of $a_0$-modulation [Eq.~(\ref{s0})], an additional factor of $2$ appears in Eq.~(\ref{sa0}). Therefore we expect different dynamics for $a_2$-modulation for a given set of interaction parameters. The results of $a_2$-modulation for $q=0$ are summarized in Fig.~\ref{fig:8} where we show the ratio of Mathieu exponents associated with the density and spin modes as a function of the driving frequency. For degenerate modes, i.e., when $\bar c_1=\bar c_0$ (dashed line in Fig.~\ref{fig:7}), $\sigma^{(0)}=2\sigma^{(+)}$, that means Faraday pattern emerges well before the spin-mixing occurs. That is also the case for $\bar c_1>\bar c_0$ (dotted-dashed line in Fig.~\ref{fig:7}). Interestingly, for $\bar c_1<\bar c_0/2$, the nature of dynamics depends on $\omega_2$. For smaller $\omega_2$,  $\sigma^{(0)}/\sigma^{(+)}<1$, i.e., spin-mixing dominates the density modulations and vice versa for large values of $\omega_2$. Making $q$ non-zero would lead to an explicit dependence of $\omega_2$ on the dynamics for any value of $\bar c_0$ and $\bar c_1$, but qualitative features remain the same.

\subsection{Modulation of both $a_0$ and $a_2$}
%
\begin{figure}
\centering
\includegraphics[width= 1.\columnwidth]{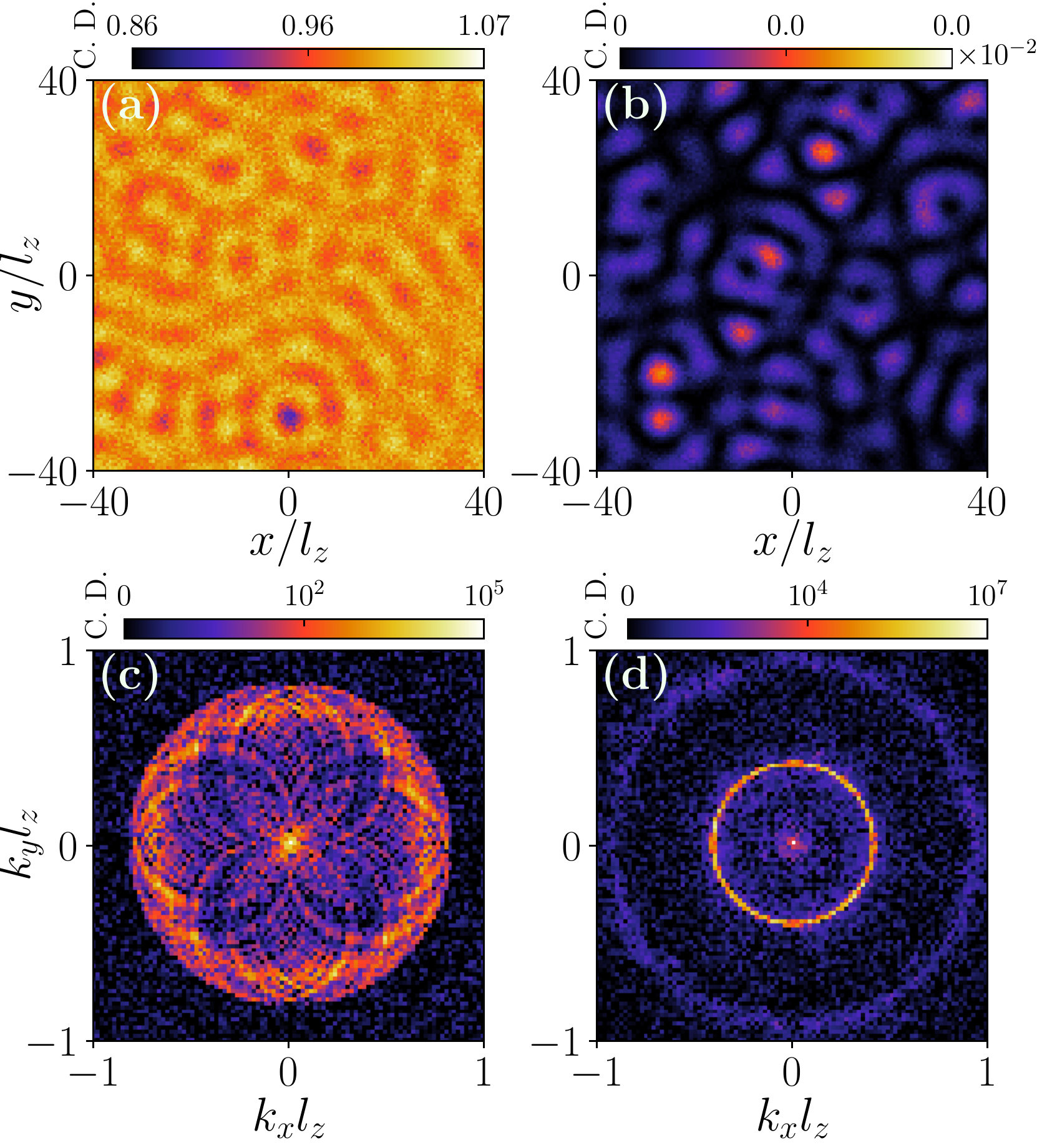}
\caption{\small{Results for simultaneous modulation of $a_0$ and $a_2$ with a phase difference of $\phi=\pi$. Real-space condensate density of (a) $m=0$ and (b) $m=\pm 1$ for $q=0$, $\bar c_0\bar n =\bar c_1\bar n=0.2\hbar \omega_z$, $\alpha_0 = 0.2$, $\alpha_2 = 0.05$, and $\omega_0/\omega_z =\omega_2/\omega_z = 0.2$ at $\omega_z t=800$. The corresponding momentum densities are shown in (c) and (d).}}
\label{fig:9} 
\end{figure}
As we have seen in the previous cases, an implicit competition exists between the density modulations (Faraday patterns) and spin-mixing. By controlling interaction strengths, quadratic Zeeman field, or the driving frequencies, we can access different scenarios in which one dominates, or both co-occur in the dynamics. Further, we show greater controllability is achieved by simultaneously modulating $a_0$ and $a_2$. To do so, we assume a phase difference $\phi$ between $a_0$ and $a_2$ modulations, i.e., $a_2(t)=\bar a_2[1+2\alpha_2\cos (2\omega_2 t+\phi)]$. Strikingly, taking $\omega_0=\omega_2$, $\alpha_2=\alpha_0 \bar g_0/\bar g_2$ and $\phi=0$ is equivalent to modulating $\bar c_0$ while keeping $\bar c_1$ constant. In that case, Eqs.~(\ref{ep1}) and (\ref{ep2}) become, 
\begin{align}
\label{epnn1}
\frac{\mathrm{d}^2 u_0 }{\mathrm{d} t^2}+\frac{1}{\hbar^2}\left[\epsilon_{{k}, 0}^2+4\bar nE_k\alpha_{0}\bar g_0\cos(2\omega_0 t)\right] u_0=0\\
 \frac{\mathrm{d}^2 u_+ }{\mathrm{d} t^2}+\frac{1}{\hbar^2}\epsilon_{{k}, \pm 1}^2 u_+=0.
 \label{epnn2}
\end{align}
The above equations convey that Faraday patterns are formed in $m=0$, but spin mixing does not occur. The numerical calculations of NLGPEs also confirm this. If we take $\phi=\pi$ and $\alpha_2= \alpha_0\bar g_0/2 \bar g_2$, $\bar c_1$ becomes periodic in time, and $\bar c_0$ remains a constant. The corresponding equations of motion are,
\begin{align}
\label{epnnn1}
\frac{\mathrm{d}^2 u_0 }{\mathrm{d} t^2}+\frac{1}{\hbar^2}\epsilon_{{k}, 0}^2 u_0=0\\
\frac{\mathrm{d}^2 u_+ }{\mathrm{d} t^2}+\frac{1}{\hbar^2}\left[\epsilon_{{k}, \pm 1}^2+2\bar n(E_k+q)\alpha_{0}\bar g_0\cos(2\omega_0 t)\right] u_+=0.
 \label{epnnn2}
\end{align}
In this case, the spin mode is unstable, leading to the population transfer from $m=0$ to $m=\pm 1$. The latter causes local depletions in the homogeneous density of $m=0$ and random density peaks emerging in $m=\pm 1$ components [see Figs.~\ref{fig:9}(a) and \ref{fig:9}(b)]. The momentum ring in the momentum density of $m=\pm 1$ [see Fig.~\ref{fig:9}(d)] arises from the spin-mixing process, $({\bm 0}, 0)+({\bm 0}, 0)\leftrightarrow({\bf k}, \pm 1)+(-{\bf k}, \mp 1)$. In contrast, the process in which a pair of atoms, each in $m=+1$ and $m=-1$ with momenta $(k_u, -k_u)$ or $(\pm k_u, \pm k_u)$ scatter into $(+{\bf k}, -{\bf k})$ or $({\bf k}, {\bf k'})$ with $|{\bf k}+{\bf k'}|= 2k_u$ populate the non-zero momenta in $m=0$ [see Fig.~\ref{fig:9}(c)] at a later time. It is starkly different from Fig.~\ref{fig:3a}(b) of the pure $a_0$ modulation where the primary unstable momenta come from the resonance with the density mode.
%
%

\subsubsection{Competing Instabilities}
\label{ci}
\begin{figure}
\centering
\includegraphics[width= 1.\columnwidth]{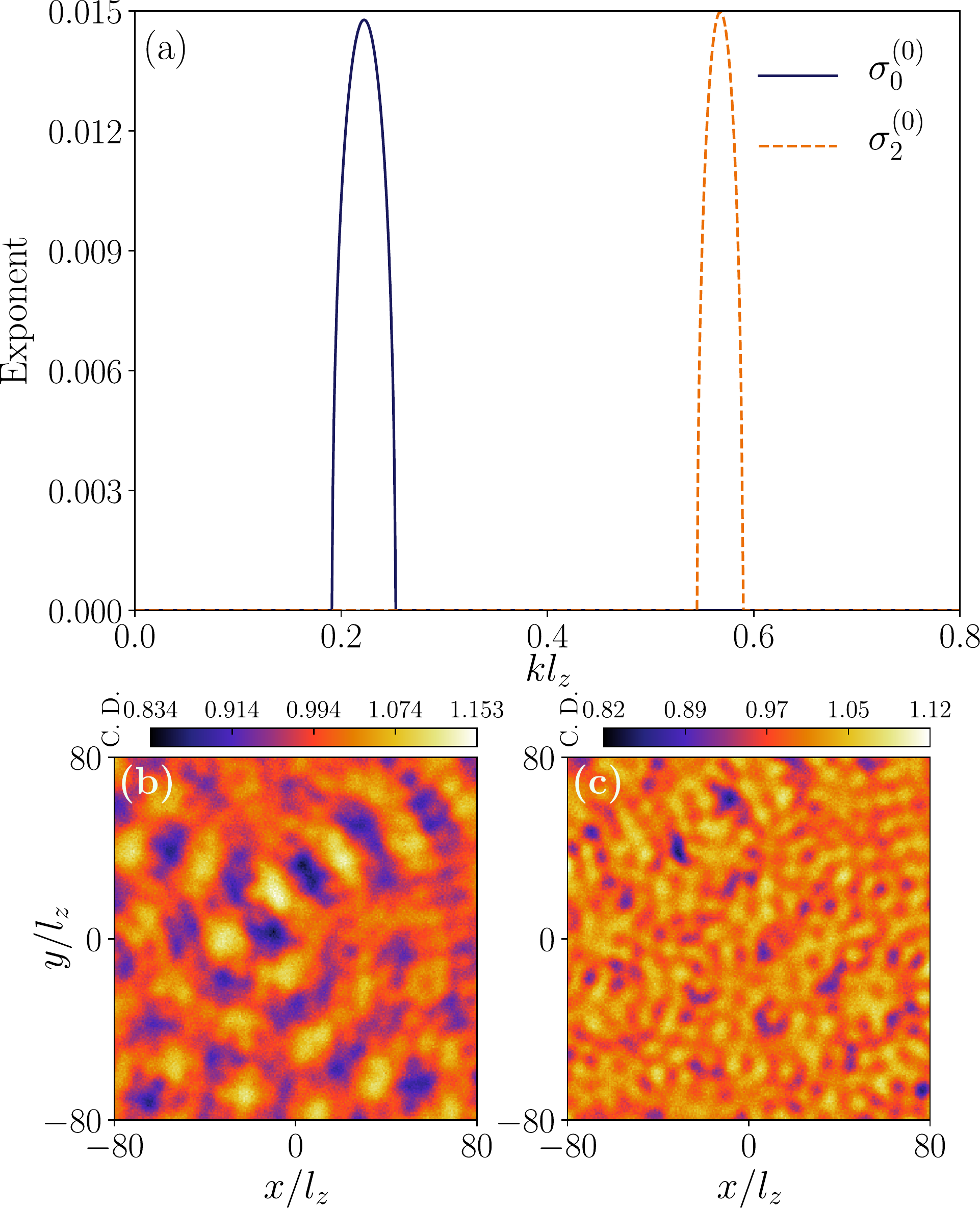}
\caption{\small{Competing instability among the density modes. (a) The two primary peaks in Mathieu exponent vs $k$ from the simultaneous modulation of $a_0$ and $a_2$ for $\bar c_0\bar n=0.2 \hbar\omega_z$, $\bar c_1\bar n=4 \hbar\omega_z$, $q=0$, $\alpha_0=0.024$, $\omega_0 = 0.1\omega_z$, $\alpha_2=0.01$, and $\omega_2 = 0.3\omega_z$. The solid line is for $a_0$, and the dashed line is for $a_2$ modulations. (b) and (c) show the density of $m=0$ component at $\omega_z t = 420$ and $\omega_z t = 460$, respectively, exhibiting the change in wavelength of patterns in time. (b) has a wavelnegth of $1/k_{u0}^{(0)}$ and (c) has a wavelength of $1/k_{u2}^{(0)}$ with $k_{u0}^{(0)}<k_{u2}^{(0)}$.}}
\label{fig:10}  
\end{figure}
%
We show that by modulating $a_0$ and $a_2$ with $\omega_0\neq\omega_2$ and carefully choosing the modulation amplitudes $\alpha_0$ and $\alpha_2$, the intriguing scenario of competing instabilities emerges. In particular, two distinct momenta of density or spin mode compete. To see the competing instability among the density modes, we take $\bar c_0\ll\bar c_1$ such that $\sigma^{(0)}\gg\sigma^{(+)}$. The primary resonances emerge from modulating $a_0$ and $a_2$ are $\epsilon_{k, 0} = \hbar \omega_0$ and $\epsilon_{k, 0} = \hbar \omega_2$ and let the corresponding unstable momenta and Mathieu exponents be ($k_{u0}^{(0)}, k_{u2}^{(0)}$) and ($\sigma_0^{(0)}, \sigma_2^{(0)}$). The relevant equation of motion is 
\begin{align}
\frac{d^2u_0}{dt^2} + \frac{1}{\hbar^2}\left[\epsilon_{k, 0}^2+\frac{4E_k\bar n}{3}\left(\alpha_0\bar g_0\cos 2\omega_0 t + 2\alpha_2\bar g_2\cos 2\omega_2 t\right)\right]u_0 =0,
\label{qpme}
\end{align} 
which is generally a quasi-periodic Mathieu equation. The stability regions of Eq.~(\ref{qpme}) studied using different approximation methods reveal a very complex structure \cite{kov18}. When $\omega_0/\omega_2$ is a rational number, Eq.~(\ref{qpme}) exhibits an overall periodicity, and the Floquet theorem becomes valid. Then, the instability regions are just a union of those arising from the independent modulations of $a_0$ and $a_2$. In Fig.~\ref{fig:10}(a), we show the two primary instability tongues associated with $a_0$ and $a_2$ modulations for $\omega_0/\omega_2=1/3$ and $\omega_2=0.3\omega_z$. The modulation amplitudes are taken such that the peak of $\sigma_0^{(0)}$ and $\sigma_2^{(0)}$ are approximately the same. In the dynamics, we observe an oscillation between the patterns of two different wavelengths $1/k_{u0}^{(0)}$ and $1/k_{u2}^{(0)}$ with $k_{u0}^{(0)}<k_{u2}^{(0)}$ [see Figs.~\ref{fig:10}(b) and \ref{fig:10}(c)]. 
\begin{figure}
\centering
\includegraphics[width= 1.\columnwidth]{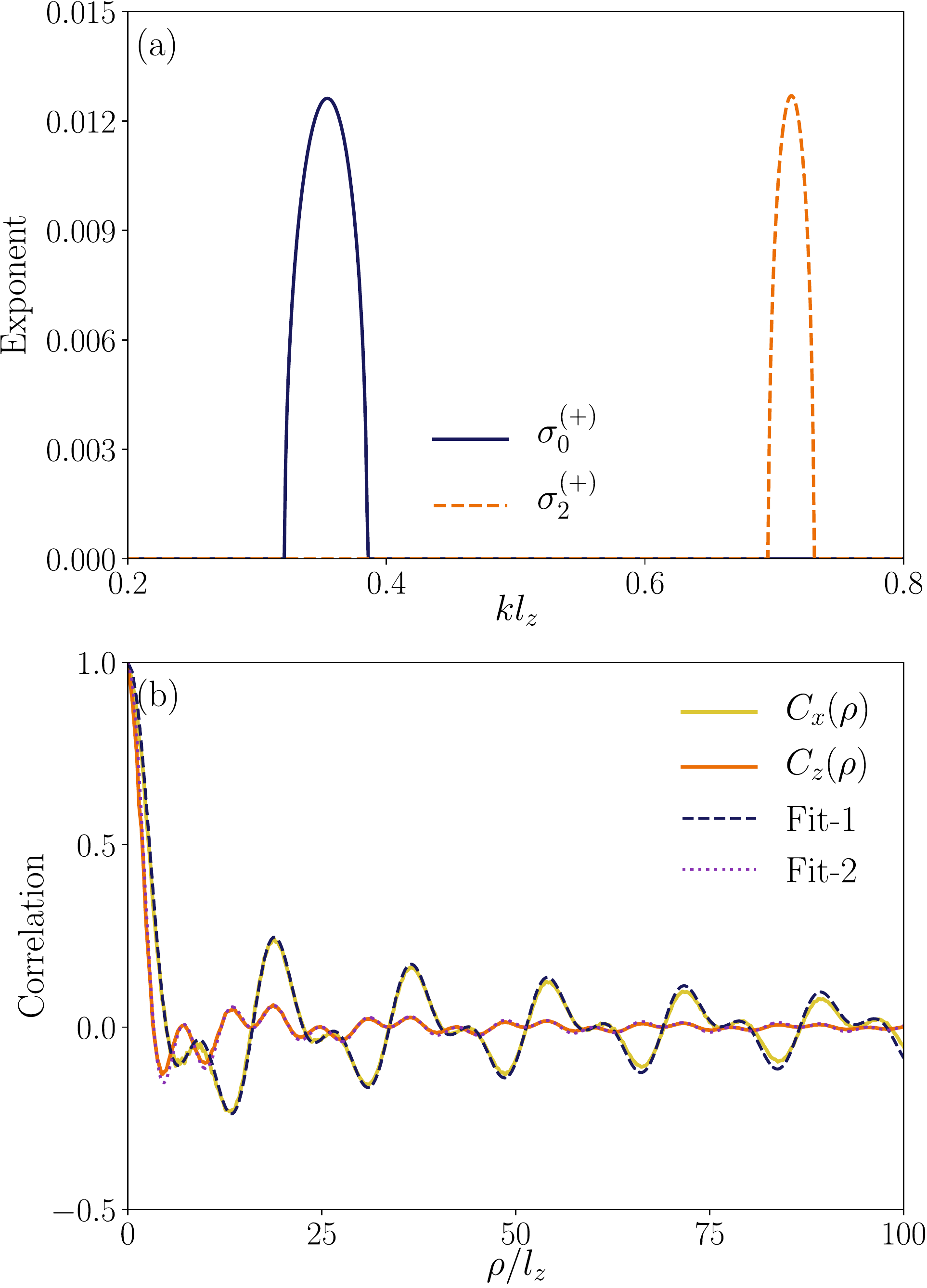}
\caption{\small{Competing instability among the spin modes. (a) The two primary peaks in Mathieu exponent vs $k$ from the simultaneous modulation of $a_0$ and $a_2$ for $\bar c_0\bar n=0.4 \hbar\omega_z$, $\bar c_1\bar n=0.05 \hbar\omega_z$, $q=0$, $\alpha_0=0.205$, $\omega_0 = 0.1\omega_z$, $\alpha_2=0.1$, and $\omega_2 = 0.3\omega_z$. The solid line is for $a_0$, and the dashed line is for $a_2$ modulations. (b) shows the spin-spin correlation functions $C_{x}(\rho)$ and $C_{z}(\rho)$. Fit-1 and Fit-2 are respectively the Eqs.~(\ref{cxy2}) and (\ref{cz2}). The numerical results in (b) are obtained by taking the average over ten realizations of noises.}}
\label{fig:11} 
\end{figure}

Similarly, for $\bar c_1\ll\bar c_0$, the competing instability arises among the two different spin-mode momenta. In that case, the relevant Mathieu equation is 
\begin{align}
\frac{d^2u_+}{dt^2} + \frac{1}{\hbar^2}\bigg[\epsilon_{k, \pm 1}^2-\frac{4(E_k+q)\bar n}{3} \nonumber \\
\left(\alpha_0\bar g_0\cos 2\omega_0 t - \alpha_2\bar g_2\cos 2\omega_2 t\right)\bigg]u_+ =0.
\end{align} 
Again, for a rational ratio of $\omega_0/\omega_2$, we obtain the instability tongues from the union of instabilities as shown in Fig.\ref{fig:11}(a). The effect of competing instabilities is directly visible in the behavior of the spin-spin correlations functions during the transient stage, and they are of the form
\begin{eqnarray}
\label{cxy2}
C_{x,y}(\rho, t) =  D(t)J_0\left(k_{u0}^{(+)}\rho\right) + \left[1-D(t)\right]J_0\left(k_{u2}^{(+)}\rho\right), \\
\label{cz2}
C_{z}(\rho, t)=J_0\left(k_{u0}^{(+)}\rho\right)J_0\left(k_{u2}^{(+)}\rho\right)
 \end{eqnarray}
where $k_{u0}^{(+)}$ and $k_{u2}^{(+)}$ are unstable momenta of the spin mode from two modulation frequencies $\omega_{0}$ and $\omega_{2}$ and $D(t)$ is the time-dependent amplitude.

\section{Experimental considerations}
\label{exp}
Now we briefly examine the experimental possibilities. As discussed above, the emergent spin textures leading to PCVs can be observed except when $\bar c_0\ll |\bar c_1|$ for which the spin mode is completely outplayed by the density mode in the instability dynamics. In the state-of-the-art experimental setups of spin-1 condensates, for instance, in $^{23}$Na, $^{87}$Rb and $^{7}$Li, the ratio $\bar c_1/\bar c_0$ is, respectively, 0.036, -0.004 \cite{kaw12} and -0.46  \cite{huh20}, which supports the formation of spin textures and PCVs. For $^{87}$Rb and $^{7}$Li, since the spin-dependent interactions are ferromagnetic, preparing the initial polar phase requires a quadratic Zeeman field. We numerically verified all three cases and confirmed the formation of spin textures and PCVs identical to the case of degenerate modes. In a given atomic setup, Feshbach resonances are required to access different regimes of interaction strengths we consider, and in particular, independent control of $a_0$ and $a_2$ is needed. The latter has been proposed to achieve via combining magnetic and rf-field-induced Feshbach resonances \cite{zha09}. Longitudinal and transverse spin-spin correlations are computed once the corresponding magnetizations are measured, as demonstrated in Ref.~\cite{seu23}.

\section{Summary and Outlook}
\label{sum}
In summary, we analyzed the density patterns and spin textures in a parametrically driven Q2D spin-1 condensate for two initial phases: ferromagnetic and polar. An initial ferromagnetic condensate is immune to periodic modulation of $a_0$, whereas, for $a_2$ modulation, it exhibits similar dynamics to that of a scalar condensate. An initial polar phase revealed interesting dynamics; for instance, a gas of polar core vortices and anti-vortices is seen with its density determined by the momentum of the unstable spin mode. Also, there is competition between Faraday patterns and spin-mixing dynamics, which can be controlled by tuning the interaction strengths, quadratic Zeeman field, or driving frequencies. When spin-mixing dynamics dominates, the spin-spin correlation functions exhibit a Bessel function behavior as a function of relative distance. Otherwise, they decay exponentially with a correlation length of the order of a spin healing length. Modulating both scattering lengths creates an exciting scenario of competing instabilities among density or spin modes. It produces the superposition of Faraday patterns or spin correlation functions of two distinct wavelengths.

Our studies open up several perspectives for future studies. For instance, one could select an appropriate initial state to engineer exotic spin textures or vortices via periodic modulation. The same analyses can be extended to condensates of higher spin where the availability of three or more scattering lengths may lead to complex scenarios. Another exciting aspect is to analyze the effect of harmonic confinement and the role of transverse excitations.

\section{Acknowledgements}

We acknowledge Chinmayee Mishra for the discussions during the initial stages of the work. We thank National Supercomputing Mission (NSM) for providing computing resources of "PARAM Brahma" at IISER Pune, which is implemented by C-DAC and supported by the Ministry of Electronics and Information Technology
(MeitY) and Department of Science and Technology (DST), Government of India. R.N. further acknowledges DST-SERB for Swarnajayanti fellowship File No. SB/SJF/2020-21/19, and the MATRICS grant (MTR/2022/000454) from SERB, Government of India and National Mission on Interdisciplinary Cyber-Physical Systems (NM-ICPS) of the Department of Science and Technology, Government of India, through the I-HUB Quantum Technology Foundation, Pune, India.

\bibliographystyle{apsrev4-1}
\bibliography{lib.bib}
\end{document}